\documentclass[acmtog,table,dvipsnames,nonacm]{acmart}

\usepackage{multirow}
\usepackage{pifont}

\usepackage[ruled]{algorithm2e} 

\SetAlFnt{\small}
\SetAlCapFnt{\small}
\SetAlCapNameFnt{\small}
\SetAlCapHSkip{0pt}

\setcopyright{none}

\begin{document}
\title{Fast and Globally Consistent Normal Orientation based on the Winding Number Normal Consistency}

\author{Siyou Lin}
\orcid{0000-0002-8906-657X}
\email{linsy21@mails.tsinghua.edu.cn}
\affiliation{%
    \institution{Department of Automation, Tsinghua University}
    \streetaddress{No. 30, Shuangqing Road}
    \city{Haidian}
    \state{Beijing}
    \postcode{100084}
    \country{China}
}

\author{Zuoqiang Shi}
\orcid{0000-0002-9122-0302}
\email{zqshi@tsinghua.edu.cn}
\affiliation{%
    \institution{Yau Mathematical Sciences Center, Tsinghua University}
    \streetaddress{No. 30, Shuangqing Road}
    \city{Haidian}
    \state{Beijing}
    \postcode{100084}
    \country{China}
}
\affiliation{%
    \institution{Yanqi Lake Beijing Institute of Mathematical Sciences and Applications}
    \streetaddress{No. 544, Hefangkou Village Huaibei Town}
    \city{Huairou}
    \state{Beijing}
    \postcode{101418}
    \country{China}
}
\authornote{Co-corresponding author}

\author{Yebin Liu}
\orcid{0000-0003-3215-0225}
\email{liuyebin@mail.tsinghua.edu.cn}
\affiliation{%
    \institution{Department of Automation, Tsinghua University}
    \streetaddress{No. 30, Shuangqing Road}
    \city{Haidian}
    \state{Beijing}
    \postcode{100084}
    \country{China}
}
\authornote{Co-corresponding author}

\begin{abstract}
Estimating consistently oriented normals for point clouds enables a number of important applications in computer graphics such as surface reconstruction. While local normal estimation is possible with simple techniques like principal component analysis (PCA), orienting these normals to be globally consistent has been a notoriously difficult problem.
Some recent methods exploit various properties of the winding number formula to achieve global consistency with state-of-the-art performance.
Despite their exciting progress, these algorithms either have high space/time complexity, or do not produce accurate and consistently oriented normals for imperfect data.
In this paper, we propose a novel property from the winding number formula, termed \textbf{Winding Number Normal Consistency (WNNC)}, to tackle this problem. The derived property is based on the simple observation that the normals (negative gradients) sampled from the winding number field should be codirectional to the normals used to compute the winding number field. Since the WNNC property itself does not resolve the inside/outside orientation ambiguity, we further propose to incorporate an objective function from Parametric Gauss Reconstruction (PGR). We propose to iteratively update normals by alternating between WNNC-based normal updates and PGR-based gradient descents, which leads to an embarrassingly simple yet effective iterative algorithm that allows fast and high-quality convergence to a globally consistent normal vector field.
Furthermore, our proposed algorithm only involves repeatedly evaluating the winding number formula and its derivatives, which can be accelerated and parallelized using a treecode-based approximation algorithm due to their special structures. Exploiting this fact, we implement a GPU-accelerated treecode-based solver. Our GPU (and even CPU) implementation can be significantly faster than the recent state-of-the-art methods for normal orientation from raw points. Our code is integrated with the popular PyTorch framework to facilitate further research into winding numbers, and is publicly available at \url{https://jsnln.github.io/wnnc/index.html}.
\end{abstract}

%
%


%
%


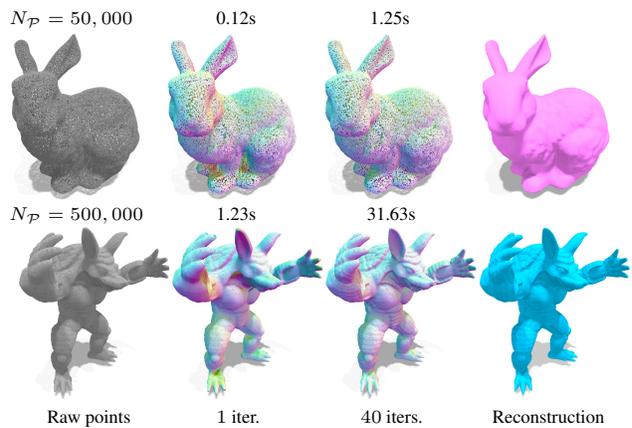
\begin{teaserfigure}
    \centering
    \includegraphics[width=\textwidth]{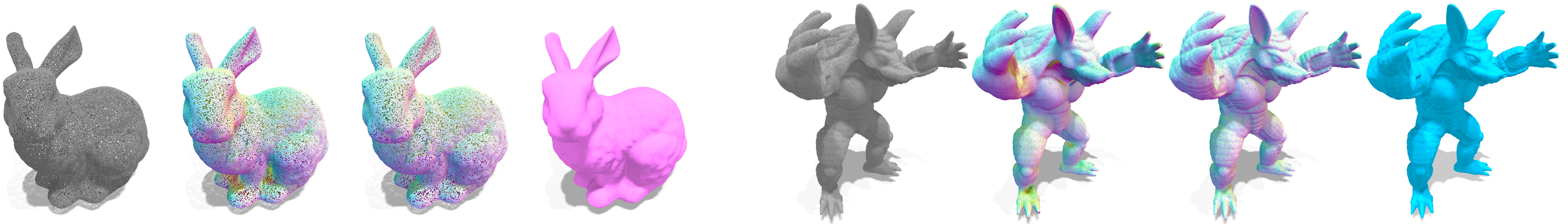}
      \begin{picture}(\textwidth,0.15cm)
      \put(0.01\textwidth,0.1cm){\small Raw points}
      \put(0.145\textwidth,0.1cm){\small $1$ iter.}
      \put(0.25\textwidth,0.1cm){\small $40$ iters.}
      \put(0.35\textwidth,0.1cm){\small Reconstruction}
      \put(0.52\textwidth,0.1cm){\small Raw points}
      \put(0.66\textwidth,0.1cm){\small $1$ iter.}
      \put(0.78\textwidth,0.1cm){\small $40$ iters.}
      \put(0.885\textwidth,0.1cm){\small Reconstruction}
      \put(0.0\textwidth,2.7cm){\small $N_{\mathcal P}=50,000$}
      \put(0.12\textwidth,2.7cm){\small 0.12s}
      \put(0.235\textwidth,2.7cm){\small 1.25s}
      \put(0.51\textwidth,3.0cm){\small $N_{\mathcal P}=500,000$}
      \put(0.66\textwidth,3.0cm){\small 1.23s}
      \put(0.785\textwidth,3.0cm){\small 31.63s}
      \end{picture}
    \caption{Given a raw point cloud, our method iteratively updates its normals to be globally consistently oriented (points are rendered as spheres colored by normals). Our iterative algorithm produces overall consistent normals even at the first iteration, and fully converges in around 40 iterations. With our treecode-accelerated implementation running on GPU~(RTX 3090), the algorithm obtains high-fidelity normals in $\sim$1 second for 50,000 points, and $\sim$30 seconds for a point cloud with half a million points.}
    \label{fig:teaser}
    \Description{Teaser.}
\end{teaserfigure}

\maketitle

\section{Introduction}

Point clouds are one of most widely used 3D representations in computer graphics applications such as geometric modeling and rendering. In most cases, consistently oriented normal vectors associated to points are necessary, e.g., for surface reconstruction~\citep{kazhdan2006pr,kazhdan2013spr,kazhdan2020sprenv,lu2018gr}. While locally estimating normals for a small patch of points can be done with simple techniques such as principal component analysis (PCA), obtaining a globally consistent orientation remains a notoriously difficult problem. Older methods mostly take a propagation-based approach by first estimating normals locally and then propagating the orientation globally~\citep{hoppe1992srup,xie2003c1,Konig2009consistent,seversky2011harmonic,metzer2021dipole}. These methods are generally fast but less robust to noise, thin structures and sharp edges where propagation easily fails. Recent developments focus more on the global consistency by introducing new optimization formulations that involve the whole point cloud~\citep{hou2022ipsr,lin2023pgr,xu2023gcno}. Despite achieving the state-of-the-art performance, these global methods often suffer from high space/time complexity, preventing their practicability.

An exciting and inspiring tread in this field is the exploration of various properties of the winding number formula (see Theorem~\ref{thm:gaussformula}). From the second line in Eq.~\eqref{eq:wnf-rhs} of the winding number formula, \citet{lin2023pgr} observed that the winding number field computed using correct normals should evaluate to $1/2$ at the input points. \citet{lin2023pgr} further turned this observation into a practical linear system, from which normals can be solved using conjugate gradients~\citep{hestenes1952cg}. Complementarily, \citet{xu2023gcno} use the first and the third line in Eq.~\eqref{eq:wnf-rhs} to define a regularization term such that the winding number evaluates to either $0$ or $1$ almost everywhere. By turning the winding number formula to optimization objectives, these two methods above successfully achieve state-of-the-art performance in estimating consistently oriented normals. However, both methods suffer from high computational complexity. The linear system in PGR~\citep{lin2023pgr} has size $N\times N$ and is dense. Running PGR~\citep{lin2023pgr} with 40,000 points consumes about 8 GB GPU memory. The high memory usage prohibits it from scaling to large point clouds. On the other hand, GCNO~\citep{xu2023gcno} has high time complexity due to its objective function being non-linear. Solving a 10,000-point model using GCNO takes about one hour.

\begin{figure}
  \centering
  \includegraphics[width=\linewidth]{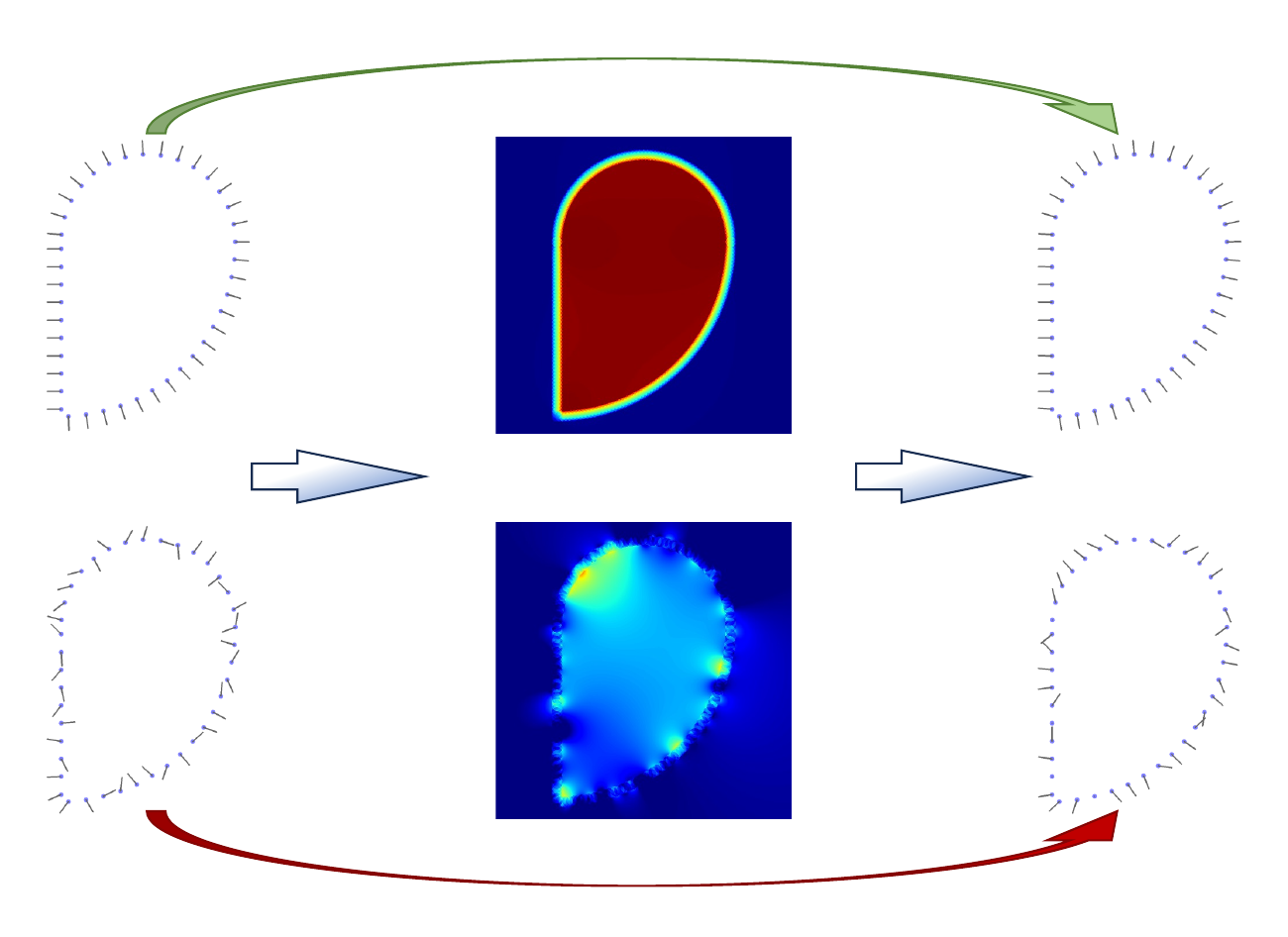}
  \begin{picture}(\linewidth,0.1cm) 
  \put(0.05\linewidth,3.35cm){\small{Normals}}
  \put(0.35\linewidth,3.35cm){\small{Winding number field}}
  \put(0.822\linewidth,3.48cm){\small{Negative}}
  \put(0.82\linewidth,3.18cm){\small{gradients}}
  \put(0.42\linewidth,6.33cm){\textcolor{Green}{\large{Aligned \ding{51}}}}
  \put(0.387\linewidth,0.32cm){\textcolor{Red}{\large{Not aligned \ding{55}}}}
  \end{picture}
  \caption{Illustration of the winding number normal consistency (WNNC) property. From left to right, we start with an input point cloud with normals, evaluate the winding number formula, and finally compute the negative field gradients at the input point positions. If the normals at the start are correctly oriented as shown in the first row, then the intermediate field is exactly the winding number field of the underlying shape. Consequently, the negative gradients at these point positions would be aligned with the input normals. We refer to this property as the winding number normal consistency (WNNC) property. Note that random normals do not satisfy the WNNC property (second row).\label{fig:wnnc-vis}}
  \Description{Illustration of the WNNC property.}
\end{figure}
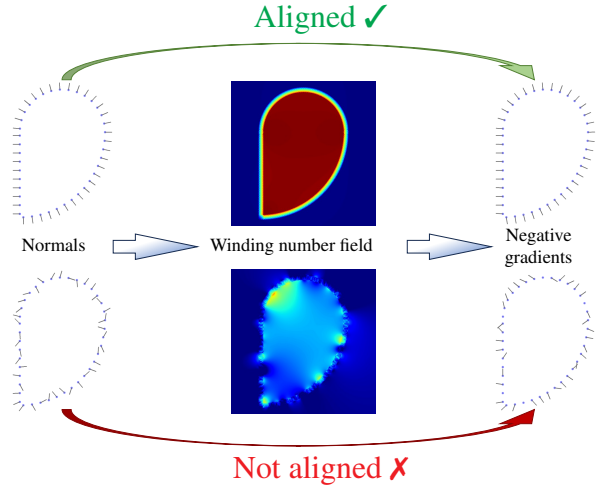

\begin{table}
    \centering
    \caption{A brief comparison of time and memory consumption for the state-of-the-art methods GCNO~\citep{xu2023gcno}, PGR~\citep{lin2023pgr}, Dipole~\citep{metzer2021dipole} and iPSR~\citep{hou2022ipsr}. The results are obtained from running an Armadillo model~\citep{krishnamurthy1996} with uniform sampling on a desktop with an Intel i9-10900X CPU (3.70GHz, 10 cores) and an NVIDIA RTX 3090 GPU. Note that the cells colored in gray are estimates based on complexity, because GCNO and PGR are too costly to run on large-scale point clouds.\label{tbl:time_teaser}}
    \small
    \begin{tabular}{c|c|c|c|c|c|c}
        \toprule
        \multirow{2}{*}{Method} & \multicolumn{2}{c}{$10^4$ points} & \multicolumn{2}{|c}{$10^5$ points} & \multicolumn{2}{|c}{$10^6$ points}\\
        \cline{2-7}
         & Time & Mem. & Time & Mem. & Time & Mem.\\
        \hline
         GCNO (CPU) & 1.1h & <1G &  \cellcolor{lightgray}{>1day} & 1.7G & \cellcolor{lightgray}{weeks} & \cellcolor{lightgray}{>10G} \\
         PGR (GPU) & 8s & 1G & \cellcolor{lightgray}{>10min} & \cellcolor{lightgray}{>30G} & \cellcolor{lightgray}{>1day} & \cellcolor{lightgray}{>3000G} \\
         Dipole (GPU) & 17s & 1G & 90s & 2.3G & 230s & 7.5G \\
         iPSR (CPU)& 18s & <1G & 50s & <1G & 380s & 2.6G \\
         \hline
         Ours (CPU) & 3s & <1G & 15s & <1G & 294s & 1G \\
         Ours (GPU) & 2s & <1G & 6s & <1G & 86s & 1.5G \\
         \bottomrule
    \end{tabular}
\end{table}

While it seems that PGR~\citep{lin2023pgr} and GCNO~\citep{xu2023gcno} have exhausted all possibilities with the winding number formula by using all three lines in Eq.~\eqref{eq:wnf-rhs} as constraints, we further exploit the winding number formula by exploring its higher-order properties. To be more specific, we propose to utilize the fact that, when using correctly oriented normals to evaluate the winding number formula, the normals (negative gradients) of the formula should be codirectional to the given normals. We refer to this property as the \textbf{winding number normal consistency (WNNC)}. An illustration of WNNC is given in Fig.~\ref{fig:wnnc-vis}. While the WNNC seems to put a strong constraint on the possible orientations of normals, unfortunately, it cannot resolve the inside/outside orientation ambiguity. We thus propose to incorporate the objective function from PGR~\citep{lin2023pgr}, which is known to encourage a consistently outward orientation but often converges at inaccurate normals. We propose to combine the advantages of both formulations by alternating between WNNC-based normal updates and PGR-based gradient descents, which leads to an embarrassingly simple yet effective iterative algorithm that allows fast and high-quality convergence to a globally consistent normal vector field. Furthermore, the proposed algorithm only involves repeatedly evaluating the winding number formula and its derivatives. Due to their special structures, these operations can be accelerated using a treecode-based algorithm~\citep{barnes1986treecode} and parallelized on GPU. Lacking an open-source implementation of the treecode algorithm that fully supports our needs (GPU-based and support for the winding number kernel derivative functions), we implement our custom treecode-based solver in CUDA. As shown in Table~\ref{tbl:time_teaser}, our implementation significantly outperforms recent state-of-the-art methods in terms of space/time complexity. Moreover, our implementation is integrated with PyTorch~\citep{paszke2019pytorch,ansel2024pytorch2} and will be made publicly available to facilitate future research into winding numbers.

We summarize our contributions as follows:
\begin{itemize}
    \item We derive the winding number normal consistency (WNNC) property from the winding number formula to provide a new mathematical formulation for estimating consistently oriented normals from raw points.
    \item We design an iterative algorithm based on the WNNC property, which achieves fast and high-quality convergence to a globally consistent normal vector field.
    \item We implement a treecode-based acceleration algorithm in CUDA integrated with PyTorch, achieving substantial improvements over prior methods in terms of computational costs. We also make our code publicly available to facilitate further research.
\end{itemize}

\section{Related Work}

The research into estimating consistently oriented normals has a long history, dating back to decades ago. In this review, we mainly introduce methods that focus on the global consistency of normal orientation. Following prior work, we categorize existing methods into three types: propagation-based, volumetric and deep learning-based. We remark that these types are not completely mutually exclusive, and we introduce existing works based on their most prominent contributions. 

\subsection{Propagation-based Methods}

Propagation-based methods mostly follow a two-step approach: They first estimate local normals with undetermined orientations, and then propagate the orientation from a seed point or local patch through its neighbors to the whole point cloud.

Local normal estimation is usually carried out by fitting a primitive geometry to local samples. Popular choices for this step include local tangent planes obtained from Principal Component Analysis (PCA)~\citep{hoppe1992srup}, implicit quadric surfaces~\citep{xie2003c1,xie2004noisydefective}, polynomials obtained by moving least squares~\citep{levin2004misi}, osculating jets~\citep{cazal2005jets,metzer2021dipole}. These techniques are often efficient and well-established.

Once unoriented normals are available, it remains to determine a sign for each of them. The general practice is to propagate local orientations through neighbors to achieve global consistency. \citet{hoppe1992srup} build a graph for input points whose edge costs are given by a measure of orientation consistency. Seeking a globally consistent orientation is then formulated as a graph optimization problem. \citet{hoppe1992srup} use the minimum spanning tree (MST) as a greedy solution, which boils down to propagating the orientation along the MST. This simple greedy approach is efficient but lacks robustness. Many follow-up methods improve one or more aspects of \citet{hoppe1992srup}. To be robust to sharp features, \citet{xie2003c1} start from multiple seed points with an improved flipping criterion to orient the whole point cloud. \citet{Konig2009consistent} further introduced Hermite curves into the flipping criterion to reduce the possibility of failures. \citet{seversky2011harmonic} use harmonic functions on the point cloud to improve the constructed MST. Going beyond MST-based approaches, other propagation schemes have been proposed as well. \citet{jakob2019parallelglobal} build a graph of points and equate a graph-edge collapse propagation with as an orientation step, and implement a fast parallel algorithm that collapse edges in a greedy manner. Instead of propagating orientations on the surface, \citet{xie2004noisydefective} first use locally oriented surfaces to determine locally oriented regions, and then propagate region orientations through a voting scheme. Recently, \citet{metzer2021dipole} propose to use neural networks to estimate consistently orientated normals for local patches, and then use dipole-based propagation to achieve global consistency.

These propagation-based methods are generally easily formulated and efficient. However, they often fail to handle difficulties such as noise, sharp edges or holes. These methods are also very sensitive to a number of factors, e.g., neighborhood size for propagation, flipping criterion, graph weight selection, etc. Thus, the applicability of these methods are limited, especially when dealing with complex geometries or imperfections.

\subsection{Volumetric Methods}

Instead of directly estimating normals for point clouds, volumetric methods adopt an indirect approach where the whole volume is partitioned into inside/outside regions. Normal orientations can be determined as the direction pointing from inside to outside.

Early works make use of space partitioning structures to identify inside/outside regions. \citet{dey2004provable} prove that, under certain assumptions, Delaunay balls of the Voronoi diagram can be well separated into outer balls and inner balls to give a desired partitioning. \citet{chen2010bot} propose the binary orientation tree which gradually tags its node corners as inside/outside while it grows.

Later mainstream volumetric methods mostly seek an implicit field $f(x)$, and define the inside region as $\{x:f(x)<c\}$ or $\{x:f(x)>c\}$. These methods usually regularize the desired implicit field with extra constraints to ensure global consistency, or directly solve for commonly used implicit fields (e.g., the signed distance field and the winding number field).

For applying additional constraints, \citet{alliez2007Voronoi} make use of unoriented normals estimated from Voronoi cells, and solve for an implicit field that minimizes an anisotropic Dirichlet energy measuring the unoriented alignment between Voronoi-based normals and the implicit field gradient. \citet{walder2005} apply value and gradient maximization together with smoothness regularization to an implicit function represented by radial basis functions (RBFs). Also based on RBFs, \citet{huang2019vipss} adopt triharmonic kernels and Duchon's energy~\citep{duchon1977} and achieve high robustness to point sparsity and non-uniform sampling.

Signed distance fields (SDFs) are a popular choice of implicit fields. \citet{mello2003inout} compute an approximate SDF on a tetrahedral grid. \citet{zhao2001} solve the Eikonal equation to obtain the SDF on a rectangular grid. Recent methods also model SDFs as neural networks and apply extra regularization as loss functions, e.g., implicit geometric regularization with the Eikonal equation~\citep{gropp2020}, or enforcing the singularity of Hessian~\citep{wang2023nsh}.

Most closely related to our work are methods based on the winding number field~\citep{barill2018fastwind,lu2018gr}. Simply speaking, the winding number field of an object equals $1$ on the inside and $0$ on the outside. Since computing the winding number field requires known normals, several methods derive properties of the winding number field and use them in turn as constraints to obtain normals. Observing the fact that the winding number field is piece-wise constant (i.e., having zero Dirichlet energy), \citet{takayama2014windingdirichlet} solve for an orientation that minimizes the Dirichlet energy of the field computed with the winding number formula. Recently, \citet{lin2023pgr} derived a linear system formulation enforcing the field to equal $1/2$ on input samples, from which normals can be efficiently solved. \citet{xu2023gcno} designed a double-well loss function encouraging the field to evaluate to either $0$ or $1$ almost everywhere. Concurrently, \citet{liu2024bim} derive a boundary energy from the harmonicity of the winding number. Optimizing the boundary energy effectively leads to consistently oriented normals.

Compared with propagation-based methods, most volumetric methods take a more global view, in the sense that their optimization objectives consider all input points throughout the whole execution process of the algorithm. As a result, these methods generally achieve much better global consistency but at a notably higher computational cost.

\subsection{Deep Learning Methods}

In the past decade, deep learning has stormed over nearly all scientific fields, with normal estimation being no exception. The powerful fitting ability of neural networks allows direct inference of normals given an input point cloud. PCPNet~\citep{guerrero2018pcpnet} and Nesti-Net~\citep{benshabat2019nestinet} use neural networks to process multiscale point features and directly infer point normals. There are also volumetric learning-based methods that infer the implicit field rather than the normals~\citep{erler2020p2s,xiao2022modified}. Different from these straight-forward regression approaches, AdaFit~\citep{zhu2021adafit} predicts point offsets prior to normal estimation to reduce the number of outliers. NeAF~\citep{li2023NeAF} further extends this idea by learning a neural angle field that offsets any given normal to the ground-truth one.

Benefiting from their ability to learn from data, deep learning methods generally exhibit better robustness to data imperfections such as noise or missing data. However, their performance strongly depends on training data and model capacity. Since current datasets are limited in quantity, complexity and variety compared to real-life objects, deep learning methods still suffer from generalizability issues.

\subsection{Other Consistent Normal Estimation Methods}

A few methods do not belong to the aforementioned categories. Shape as Points (SAP)~\citep{peng2021dpsr} derives a differentiable version of Poisson Surface Reconstruction (PSR)~\citep{kazhdan2006pr}, and uses it to update normals such that the reconstructed mesh has a low error with the input points. Since the differentiable PSR~\citep{peng2021dpsr} is performed in the frequency domain with fast Fourier transform, its reconstruction accuracy is limited by grid resolution. Iterative Poisson Surface Reconstruction (iPSR)~\citep{hou2022ipsr} starts with applying Screened Poisson Surface Reconstruction (SPSR)~\citep{kazhdan2013spr} to points with arbitrary normals, and iteratively resample normals from the mesh reconstructed in the previous iteration. \citet{hou2022ipsr} show that this simple iterative procedure can update normals into a consistent orientation in only 5$\sim$30 of steps.
\section{Preliminaries}

Our method builds on the winding number formula~\citep{barill2018fastwind} (also referred to as the Gauss formula in potential theory~\cite{lu2018gr}), and is closely related to PGR~\citep{lin2023pgr}. We begin with a brief review of them.

\begin{theorem}[Winding Number]
Let $\Omega\subset\mathbb{R}^3$ be an open and bounded region. Suppose its boundary $\partial\Omega$ is smooth and $n(y)$ is the outward unit normal at $y\in\partial\Omega$. Then for any $y=(y_1,y_2,y_3)\in\mathbb{R}^3$,
\begin{eqnarray}
F(y)&=& \int_{\partial\Omega}((\nabla\Phi)(y-x))\cdot n(x)\,\mathrm{d}S(x)\label{eq:wnf-lhs}\\
    &=& \left\{\begin{array}{lll}0 &,& y\in\bar\Omega^c\\
                         1/2 &,& y\in\partial\Omega\\
                         1 &,& y\in\Omega\end{array}\right.\label{eq:wnf-rhs}
\end{eqnarray}
is the indicator function of $\Omega$. Here, $\Phi(y)$ is the fundamental solution to the $3$-dimensional Laplace equation:
\begin{equation}
    \Phi(y)=\frac{1}{4\pi|y|},\quad\nabla\Phi(y)=-\frac{y}{4\pi|y|^3}.
\end{equation}
\label{thm:gaussformula}
\end{theorem}

With a known normal vector field $n(x)$, the indicator function can be directly evaluated by Theorem~\ref{thm:gaussformula}. To numerically evaluate the winding number given a point set $\mathcal P=\{x_j\}_{j=1}^{N_{\mathcal P}}$ that samples the surface $\partial\Omega$, \citet{lin2023pgr} adopt the following simple discretization:
\begin{eqnarray}
F(y)&=& \int_{\partial\Omega}((\nabla\Phi)(y-x))\cdot n(x)\,\mathrm{d}S(x)\nonumber\\
    &\approx& \sum_{j=1}^{N_{\mathcal P}}\underbrace{\frac{-(y-x_j)}{4\pi|y-x_j|^3}}_{=((\nabla\Phi)(y-x_j))\in\mathbb{R}^3}\cdot\underbrace{n(x_j)\,\sigma(x_j)}_{=\mu_j\in\mathbb{R}^3}\\
    &=&\sum_{j=1}^{N_{\mathcal P}}((\nabla\Phi)(y-x_j))\cdot\mu_j.\label{eq:wnf-discretization}
\end{eqnarray}
Here, $\sigma(x_j)$ denotes the local surface element area, $n(x_j)$ denotes the unit vector and $\mu_j$ denotes their product, i.e., the oriented surface element at $x_j$. \citet{lin2023pgr} further noticed that \eqref{eq:wnf-rhs} in fact defines a linear equation in $\mu_j$:
\begin{equation}
\sum_{j=1}^{N_{\mathcal P}}(\nabla\Phi)(x_i-x_j)\cdot\mu_j=1/2,\ \text{for all }x_i\in\mathcal{P}.\label{eq:pgr-eq}
\end{equation}
\citet{lin2023pgr} then rewrite Eq.~\eqref{eq:pgr-eq} into a regularized symmetric system, which can then be efficiently solved with conjugate gradients~\citep{hestenes1952cg}. Finally, the $\mu_j$'s are inserted back into Eq.~\eqref{eq:wnf-discretization} to compute the winding number field.

Our method also applies the discretization in Eq.~\eqref{eq:wnf-discretization} as the representation of an indicator function, and we also utilize the constraints Eq.~\eqref{eq:pgr-eq} for obtaining the $\mu_j$'s. In addition, we propose an iterative scheme which incorporates the WNNC that is absent from previous work, together with a GPU-based treecode-accelerated implementation to improve the efficiency of our algorithm, detailed in Section~\ref{sec:method}.



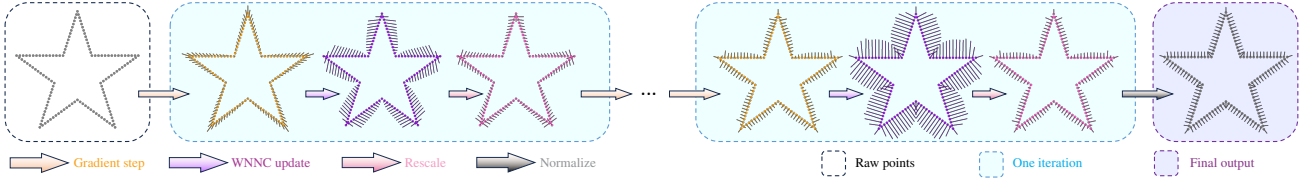
\begin{figure*}
  \centering
  \includegraphics[width=\linewidth]{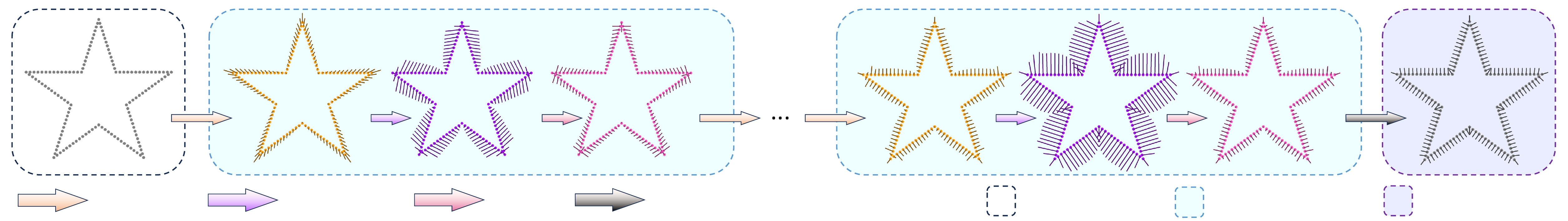}
  \begin{picture}(\linewidth,0.0cm) 
  \put(0.06\linewidth,0.55cm){\scriptsize{\textcolor{YellowOrange}{Gradient step}}}
  \put(0.18\linewidth,0.55cm){\scriptsize{\textcolor{Mulberry}{WNNC update}}}
  \put(0.312\linewidth,0.55cm){\scriptsize{\textcolor{Lavender}{Rescale}}}
  \put(0.415\linewidth,0.55cm){\scriptsize{\textcolor{Gray}{Normalize}}}
  \put(0.655\linewidth,0.55cm){\scriptsize{Raw points}}
  \put(0.775\linewidth,0.55cm){\scriptsize{\textcolor{Cyan}{One iteration}}}
  \put(0.91\linewidth,0.55cm){\scriptsize{\textcolor{Fuchsia}{Final output}}}
  \end{picture}
  \caption{The overall pipeline of our iterative algorithm. Each iteration consists of three steps: a gradient step w.r.t. a quadratic energy function proposed in \citet{lin2023pgr}, a normal update step based on the WNNC property, and a rescaling step to ensure numerical stability. Our algorithm can generally produce an overall correct orientation at the first iteration, and gradually converges to details in a few dozens of iterations. \label{fig:pipeline}}
  \Description{Illustration of the pipeline of our iterative algorithm.}
\end{figure*}

\section{Method}\label{sec:method}

\subsection{Notations}

Throughout the paper, we use $\mathcal P=\{x_j\}_{j=1}^{N_{\mathcal P}}$ to denote the unoriented input points, which are assumed to distribute on or near the surface of a 3D shape. Our goal is to estimate normals\footnote{Strictly speaking, $\mu_j$ are oriented surface elements with their lengths being surface element areas. However, in most cases it is only the orientation that matters. We thus use the word \textit{normals} to refer to $\mu_j$ for simplicity.} $\mu_j$ that are consistently oriented. For the remaining part of this paper, we only use the discretized winding number formula Eq.~\eqref{eq:pgr-eq}. We also rewrite it as $F(y;\mu)$ to highlight the dependency of $F$ on $\mu$. Furthermore, we consider $\mu$ as unknowns. Note that $F(y;\mu)$ does not represent the actual winding number unless $\mu$ represents the actual normal vector field, but we call $F(y;\mu)$ the winding number formula nonetheless for convenience. The rest of this paper is devoted to finding a solution $\mu$ that represents consistently oriented normals. 

\subsection{Winding Number Normal Consistency}

In this section, we introduce our core contribution, the winding number normal consistency (WNNC). The WNNC property is based on a simple observation: If the given normals $\mu_j$ are consistently outward-pointing, i.e., satisfying the assumptions in Theorem~\ref{thm:gaussformula}, then the winding number field $F(y;\mu)$ computed using these normals should be the indicator field of the corresponding shape. Consequently, the negative gradient of $F(y;\mu)$ at $x_i$ should point to the same direction as $\mu_i$. Fig.~\ref{fig:wnnc-vis} illustrates two cases, where one starts with (possibly incorrectly oriented) normals $\mu$, evaluates $F(y;\mu)$ and then resamples its negative gradient at each $x_i$. Note that if the original normals $\mu$ are correctly oriented, the resampled ones would still be aligned with them. The WNNC property can be expressed mathematically as follows: If $\mu_j$ are consistently outward-pointing, then
\begin{equation}
\mu_i=\lambda_i(-\nabla F(x_i;\mu))\quad\textrm{for some}\quad\lambda_i>0.\label{eq:wnnc-property-v1}
\end{equation}
Fortunately, it is not necessary to compute $F(y;\mu)$ first and then its gradient. The discretized expression of $\nabla F(y;\mu)$ can be directly derived from Eq.~\eqref{eq:wnf-discretization}:
\begin{equation}
    \nabla F(y;\mu)=\nabla\Big(\sum_j((\nabla\Phi)(y-x_j))\cdot\mu_j\Big)=\sum_j((H\Phi)(y-x_j))\cdot\mu_j,
\end{equation}
where $H\Phi$ is the Hessian of $\Phi$:
\begin{equation}
    H\Phi(y)=-\frac{I_3}{4\pi|y|^3}+\frac{3}{4\pi|y|^5}\left[\begin{matrix}y_1\\y_2\\y_3\end{matrix}\right][y_1\ y_2\ y_3].
\end{equation}
To simplify notations, we define the mapping $\mu\mapsto\hat\mu$, where $\hat\mu_i:=-\nabla F(x_i;\mu)$, as $\hat\mu=G(\mu)$. In other words, $G$ takes normals $\mu$ as input, and evaluates the negative gradients of the winding number field at the same point locations. Note that the expression of $G$ depends only on the point locations $x_j\in\mathcal P$.

Finally, we can conclude the WNNC property in a simple mathematical form: If $\mu_j$ are consistently outward-pointing, then
\begin{equation}
    \mu_{i}=\lambda_i(-\nabla F(y_i;\mu))=\lambda_iG(\mu)_i\quad\textrm{for some}\quad\lambda_i>0.\label{eq:wnnc-property-v2}
\end{equation}

\subsection{WNNC-based Iterative Algorithm}\label{subsec:iterative-algorithm}

In this section, we describe our algorithm that utilizes the WNNC property Eq.~\eqref{eq:wnnc-property-v2} to achieve fast and high-quality convergence. At first sight, one may think it is natural to define an objective function to minimize the discrepancy between the directions of $\mu$ and $G(\mu)$, e.g., using the cosine similarity between them. However, this would lead to a non-linear problem that may be complicated to solve. We observe that the WNNC condition Eq.~\eqref{eq:wnnc-property-v2} seems to hint $\mu$ is the \textit{fixed point} of the mapping $\mu\mapsto G(\mu)$ (up to per-point scaling). Inspired by fixed-point iteration algorithms in classical numerical analysis, we attempt to adopt an iterative scheme for solving normals using the WNNC property, where we update $\mu$ using $\mu\mapsto G(\mu)$. We call $\mu\mapsto G(\mu)$ a WNNC update.

Despite the similarity between WNNC updates and fixed-point iterations, merely applying the fixed-point iteration algorithm is infeasible due to the following reasons. (a) A trivial solution $\mu=0$ is clearly a fixed point of $\mu\mapsto G(\mu)$. (b) Both consistently outward-pointing normals and consistently inward-pointing normals are invariant under WNNC updates (up to scaling), which suggests the WNNC alone may not be able to flip incorrect normals. In practice, we also find merely using WNNC updates is insufficient for convergence (see Sec.~\ref{sec:wnncalone}). (c) The WNNC property involves only the normal directions, but says nothing about the scale changes $\lambda_i$. Thus, $\mu$ is not a fixed point of $\mu\mapsto G(\mu)$ in a traditional sense. Furthermore, we empirically found $\lambda_i$ to range from $10^4$ to $10^{10}$ for typical datasets, which would blow up numerically in only a few iterations.

Since WNNC updates alone are insufficient, we propose to incorporate the constraints Eq.~\eqref{eq:pgr-eq} from PGR~\citep{lin2023pgr} as an additional objective function:
\begin{equation}
    E=\sum_i\Big|\sum_{j=1}^{N_{\mathcal P}}(\nabla\Phi)(x_i-x_j)\cdot\mu_j-1/2\Big|^2,
\end{equation}
We remark that, as discussed in \citet{lin2023pgr}, the above equation can resolve the inside/outside ambiguity and encourages the resulting normals to be consistently outward-pointing.
Again, to simplify notations, we define the mapping $\mu\mapsto s$, where $s_i:=F(x_i;\mu)$, as $s=A(\mu)$. In other words, $\mu\mapsto s=A(\mu)$ evaluates the winding number field at each $x_i$ with given normals $\mu$. In this way, we can rewrite $E$ as
\begin{equation}
    E=E(\mu;A,b)=\|A(\mu)-b\|^2,\quad\textrm{where}\quad b=(1/2,\cdots,1/2)^T.\label{eq:value-energy}
\end{equation}

Our iterative algorithm aims to update normals using both WNNC updates $\mu\mapsto G(\mu)$ and $E(\mu;A,b)$. More specifically, our algorithm starts with an all-zero initial vector field $\mu=0$. During each iteration, we first take a gradient step w.r.t. $E(\mu;A,b)$, and then update the normals using an WNNC update. To solve the scale change problem during WNNC updates mentioned above, we rescale each $\mu_i$ to have the same length as before each WNNC update. The algorithm is summarized as Algorithm~\ref{alg:wnncalg-nosmooth}, and Fig.~\ref{fig:pipeline} presents a graphical overview of our iterative pipeline.
\begin{algorithm}
\SetAlgoLined
\KwData{Unoriented point set $\mathcal P=\{x_j\}$ with $N_{\mathcal P}$ points.}
\KwResult{Normals $\mu$.}
$\mu\gets0\in\mathbb{R}^{N\times3}$\;
\For{$i\gets1$ \KwTo $n$}{
    $\mu\gets{\rm grad\_step}(\mu;A,b)$\tcp*[r]{as in Alg.~\ref{alg:gradstep}}
    $\hat\mu\gets{G(\mu)}$\tcp*[r]{WNNC update}
    $\mu\gets{\rm rescale}(\mu,\hat\mu)$\tcp*[r]{$\mu_i\gets\hat\mu_i|\mu_i|/|\hat\mu_i|$}
}
 \caption{WNNC-based iterative algorithm (no smoothing)}
 \label{alg:wnncalg-nosmooth}
\end{algorithm}

Note that $\mu\mapsto A(\mu)$ is a linear operator, and thus $E(\mu;A,b)$ is a quadratic form. A single gradient step of $\mu$ w.r.t. $E(\mu;A,b)$ is presented in Algorithm~\ref{alg:gradstep}. This can be derived with basic linear algebra.
\begin{algorithm}
\SetAlgoLined
\KwData{Matrix $A$ and vectors $\mu,b$.}
\KwResult{Updated $\mu$ after a single gradient step w.r.t. $\|A\mu-b\|^2$.}
$r\gets A^Tb-A^TA\mu$\;
$\alpha\gets r^Tr/r^TA^TAr$\;
$\mu\gets\mu+\alpha r$\;
 \caption{Single-step gradient update grad\_step($\mu;A,b$).}
 \label{alg:gradstep}
\end{algorithm}

\subsection{Resolving Singularities}\label{subsec:width}

The steps in Algorithm~\ref{alg:wnncalg-nosmooth} are none other than evaluating expressions of the form $(\nabla\Phi)(x_i-x_j)$ and $(H\Phi)(x_i-x_j)$. However, both expressions become singular when $x_i$ approaches $x_j$. Using their raw forms inevitably brings numerical instability. Following previous work~\citep{lu2018gr}, we set their values to $0$ if $\|x_i-x_j\|<w$, where $w>0$ is a constant called the \textit{smoothing width}. This change affects the operators $A$ and $G$. We denote their modified versions by $A_w$ and $G_w$, respectively. As reported in previous work~\citep{lu2018gr,lin2023pgr}, a small $w$ retains more details but is less robust to noise and other imperfections, while a large $w$ leads to an overly smooth winding number field. In practice, we find that using a fixed width as in PGR~\citep{lin2023pgr} either leads to oversmoothing or cannot ensure global consistency. We thus propose a width scheduling scheme, where we start with a large $w=w_2$ and linearly decreases it to a small $w=w_1$ as the algorithm proceeds. In practice, this allows the iterative algorithm to remain stable during early iterations, and to enhance details in late iterations. The final algorithm is described in Algorithm~\ref{alg:wnncalg-final}.

\begin{algorithm}
\SetAlgoLined
\KwData{Unoriented point set $\mathcal P=\{x_i\}$ with $N_{\mathcal P}$ points, $w_1\leq w_2$.}
\KwResult{Normals $\mu$.}
$\mu\gets0\in\mathbb{R}^{N\times3}$\;
\For{$i\gets1$ \KwTo $n$}{
    $w=w_2\frac{n-i}{n-1}+w_1\frac{i-1}{n-1}$\;
    $\mu\gets{\rm grad\_step}(\mu;A_w,b)$\tcp*[r]{as in Alg.~\ref{alg:gradstep}}
    $\hat\mu\gets{G_w(\mu)}$\tcp*[r]{WNNC update}
    $\mu\gets{\rm rescale}(\mu,\hat\mu)$\tcp*[r]{$\mu_i\gets\hat\mu_i|\mu_i|/|\hat\mu_i|$}
}
 \caption{WNNC-based iterative algorithm (with smoothing)}
 \label{alg:wnncalg-final}
\end{algorithm}

From now on we assume the smoothing width is always used, but for convenience we may not make it explicit in notations.

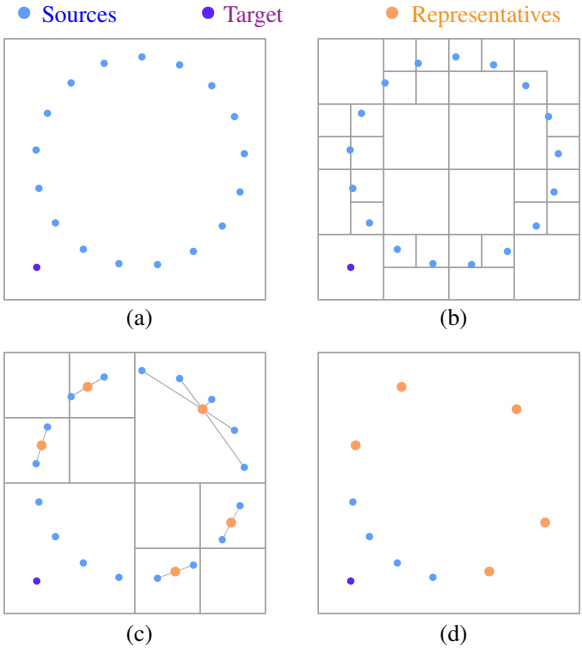
\begin{figure}
  \centering
  \includegraphics[width=\linewidth]{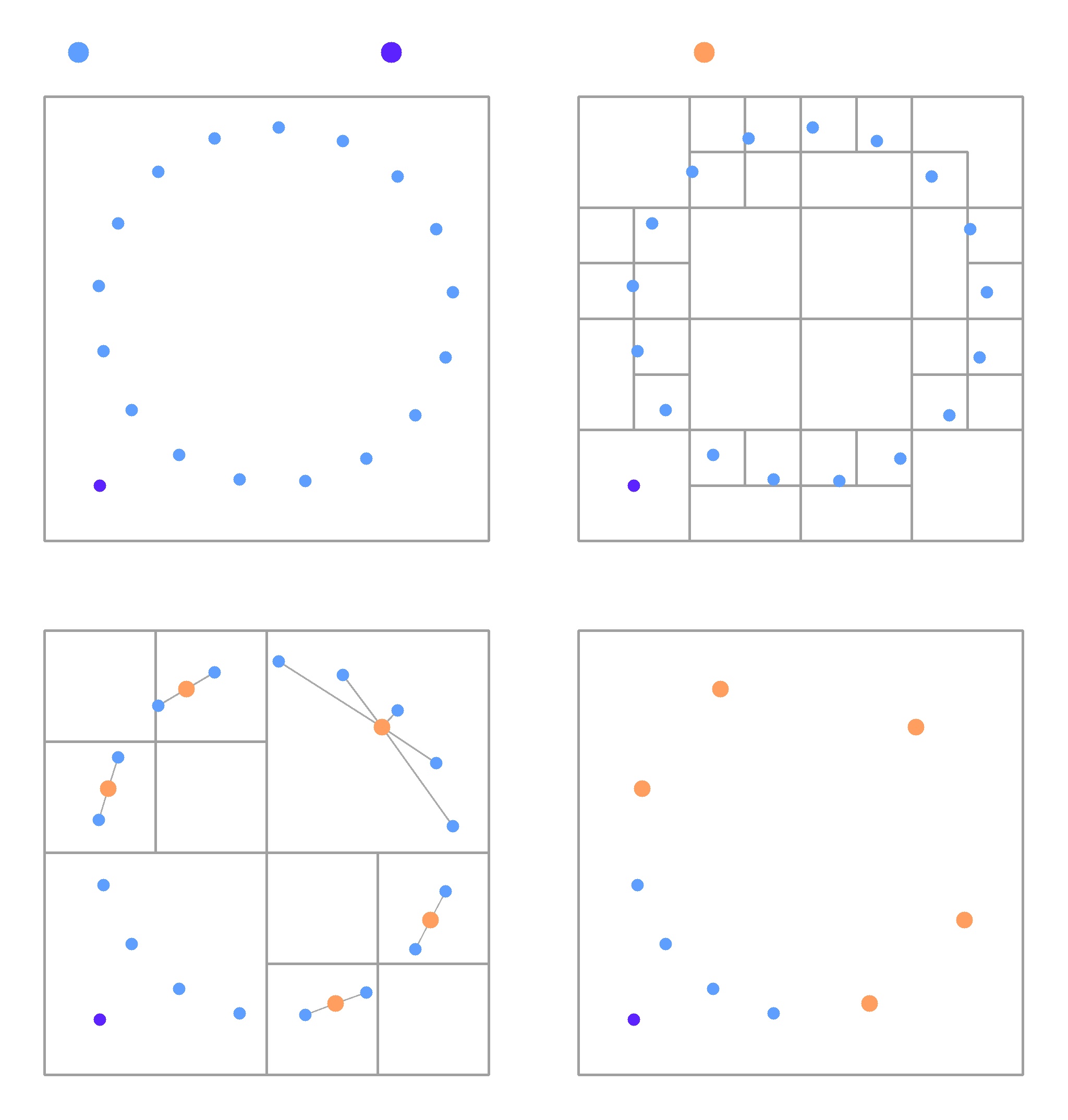}
  \begin{picture}(\linewidth,0.0cm)
  \put(0.235\linewidth,4.7cm){\small (a)}
  \put(0.735\linewidth,4.7cm){\small (b)}
  \put(0.235\linewidth,0.4cm){\small (c)}
  \put(0.735\linewidth,0.4cm){\small (d)}

  \put(0.10\linewidth,8.84cm){\small \textcolor{blue}{Sources}}
  \put(0.39\linewidth,8.84cm){\small \textcolor{Plum}{Target}}
  \put(0.69\linewidth,8.84cm){\small \textcolor{BurntOrange}{Representatives}}
  \end{picture}
  \caption{Illustration of the treecode-based approximation algorithm~\citep{barnes1986treecode}. (a) We wish to sum up interactions between multiple \textcolor{blue}{source points} and the \textcolor{Plum}{target point}. (b) The space is partitioned by an octree hosting the source points, and a representative point is computed in each node (not shown here). (c) Based on the proximity between the target point and the octree nodes, \textcolor{BurntOrange}{representative points} are chosen. Attributes of the source points are aggregated to their corresponding representative points. (d) For far-field interactions, representative points are used instead of the original source points.\label{fig:treecode}}
\end{figure}

\subsection{Treecode-based Acceleration on GPU}

As described in Algorithm~\ref{alg:gradstep} and Algorithm~\ref{alg:wnncalg-final}, our iterative scheme mainly involves evaluating the three linear operators $\mu\mapsto A(\mu)$, $p\mapsto A^T(p)$ and $\mu\mapsto G(\mu)$. A naive method for evaluating each of these operators has time complexity $O(N_{\mathcal P}^2)$. The computational time becomes prohibitively long as the number of input points grows.

The idea for accelerating these operators is based on the following observation: Each operator can be seen as summing up the \textit{interactions} between $x_i$ (targets) and $x_j$ (sources) through kernel functions $\nabla\Phi(x_i-x_j)$ or $\nabla H(x_i-x_j)$, and since both kernels vanish sufficiently fast as $|x_i-x_j|$ increases, far-field interactions of a group of neighboring sources can be well approximated by a new representative source (see Fig.~\ref{fig:treecode}).

To build representative sources, we first build an octree $\mathcal T$ to spatially partition the input point set. The partitioning stops if the node contains only one point or if the user-specified maximum depth $D$ is reached. Then, at each iteration, each octree node accumulates point attributes (such as $\mu_i$) from the points it contains and select a representative point as a weighted average of all points in the node.
The representative point of each node is determined as follows. Let $B$ be some node of the octree and let $\mathcal P_{B}$ be the subset of input points that lie in $B$. Let $\nu_j$ be an attribute vector of $x_j$ that we wish to propagate to $x_i$ through an interaction (e.g., $\nu_j=\mu_j$ when evaluating $\mu\mapsto A(\mu)$ or $\mu\mapsto G(\mu)$, and $\nu_j=s_j$ when evaluating $s\mapsto A^Ts$). The $\nu$-representative of the node $B$ is located at
\begin{equation}
    x_{B,\nu}=\sum_{x_i\in\mathcal P_{B}}\frac{|\nu_i|}{\sum_{x_j\in\mathcal P_{B}}|\nu_j|}x_i,\label{eq:node-rep-loc}
\end{equation}
associated with the representative attribute vector
\begin{equation}
    \nu_{B}=\sum_{x_i\in\mathcal P_{B}}\nu_i.\label{eq:node-rep-vec}
\end{equation}
With the $\nu$-representatives already obtained, the operators above can be evaluated by simple tree traversal. Algorithm~\ref{alg:treecode-single} describes the process in detail using $\mu\mapsto s=A(\mu)$ as an example.
\begin{algorithm}
\SetAlgoLined
\KwData{Query point $x_i$; normals at the current iteration $\mu=(\mu_i)\in\mathbb{R}^{N_{\mathcal P}\times3}$; octree $\mathcal T$ with $\mu$-representatives already computed; the starting tree node $B$.}
\KwResult{An output scalar $s_i$ associated to $x_i\in\mathcal P$.}
$s_i\gets0$\;
\uIf{$|x_i-x_{B,\mu}|>c\cdot({\rm the\ width\ of\ }B)$}{
    $s_i\gets s_i+(\nabla\Phi)(x_i-x_{B,\mu})\cdot\mu_B$\tcp*[r]{modified by $w$}
}
\uElseIf{$B$ is not a leaf}{
    \For{each child node $B_{\rm c}$ of $B$}{
        $s_i\gets s_i+{\rm apply\_A}(x_i,\mu,B_{\rm c})$\;
    }
}
\Else{
    \For{each point $x_j\in\mathcal P_B$}{
        $s_i\gets s_i+(\nabla\Phi)(x_i-x_j)\cdot\mu_j$\tcp*[r]{modified by $w$}
    }
}
\Return $s_i$\;
\caption{${\rm apply\_A}(x_i,\mu,B)$.}
\label{alg:treecode-single}
\end{algorithm}
In Algorithm~\ref{alg:treecode-single}, $B$ is a node of the octree $\mathcal T$ with its $\mu$-representatives already build, and ${\rm apply\_A}(B)$ is a function that computes
\begin{equation}
    \sum_{x_j\ {\rm inside}\ B}(\nabla\Phi)(x_i-x_j)\cdot\mu_j.
\end{equation}
To get $s_i$ in $s=A(\mu)$, simply call ${\rm apply\_A}(x_i,\mathcal T_{\rm root})$. Note that $c$ and $w$ are hyperparameters. $c$ determines whether to use the representative as an approximation, and $w$ is the smoothing width in Sec.~\ref{subsec:width}. Finally, we parallelize all evaluations ${\rm apply\_A}(x_i,\mathcal T_{\rm root})$ across all $x_i$. Other operators $A^T$ and $G$ are accelerated in the same way.

We remark that the idea of using a tree structure to approximate far-field interactions dates back to \citet{barnes1986treecode}, which was originally proposed to accelerate the gravitational 
$N$-body problem. The tree-based algorithm is now referred to as treecode~\citep{yokota2011}. The fast multipole method (FMM)~\citep{greengard1997fmm} pushes this idea further by using representatives for target points as well and by using expansions for propagation kernels for precision control. Previous works~\citep{barill2018fastwind,lu2018gr} have also utilized these acceleration techniques to accelerate evaluating winding numbers. While treecodes and FMM are well-studied with publicly available libraries, to the best of our knowledge, there is currently no open-source implementation of treecode or FMM that suits our requirements: (a) GPU-based; (b) support for the operators used in our algorithm: $A,A^T$ and $G$. For this reason, we implement our own CUDA kernels and integrate them in PyTorch~\citep{paszke2019pytorch,ansel2024pytorch2}. Note that the $A^T$ operator is exactly the backward function for $A$, which means our code also supports differentiable programming involving winding numbers. We hope our implementation will further advance the research into winding numbers.

\section{Experiments}

\subsection{Experiment Setup}

\subsubsection{Implementation Details}

In our implementation, the input point cloud is normalized to fit into the cube $[-1,1]^3$ with a margin of $1/11$ from the cube boundaries. The smoothing widths are set as $w_1=0.002$ and $w_2=0.016$ unless otherwise stated. In Algorithm~\ref{alg:treecode-single}, the maximum tree depth is $D=15$, and the threshold for using a representative source is $c=2$. We run 40 iterations for each model.

\subsubsection{Baselines}

We choose several recent state-of-the-art methods, dipole propagation~\citep{metzer2021dipole}, iterative Poisson Surface Reconstruction (iPSR)~\citep{hou2022ipsr}, Parametric Gauss Reconstruction (PGR)~\citep{lin2023pgr} and GCNO~\citep{xu2023gcno}, as the baselines. We use their default parameters for our experiments unless otherwise mentioned. For Dipole~\citep{metzer2021dipole}, we use \verb|orient_large| for point clouds with more than 50,000 points, and \verb|orient_pointcloud| otherwise. All experiments are done on a desktop computer with an Intel i9-10900X CPU (3.70GHz, 10 cores) and an NVIDIA RTX 3090 GPU.

\begin{figure*}
  \centering
  \includegraphics[width=\textwidth]{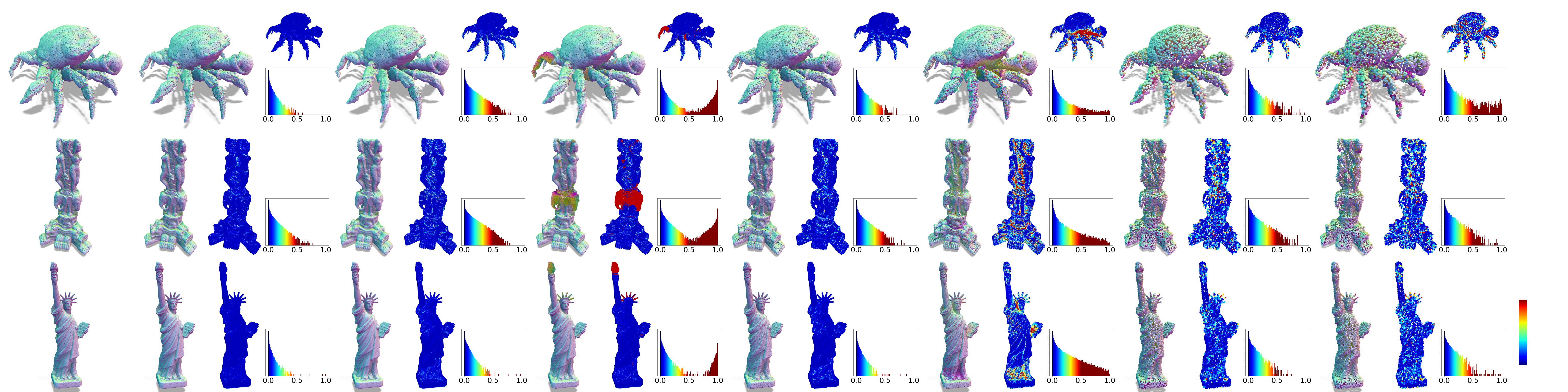}
  \begin{picture}(\textwidth,0.5cm)
  \put(0.032\textwidth,0.2cm){\small GT}
  \put(0.120\textwidth,0.2cm){\small Ours}
  \put(0.240\textwidth,0.2cm){\small iPSR}
  \put(0.360\textwidth,0.2cm){\small Dipole}
  \put(0.490\textwidth,0.2cm){\small Ours$^\dag$}
  \put(0.620\textwidth,0.2cm){\small PGR$^\dag$}
  \put(0.740\textwidth,0.2cm){\small Ours$^\ddag$}
  \put(0.860\textwidth,0.2cm){\small GCNO$^\ddag$}
  \put(0.968\textwidth,0.6cm){\small 0}
  \put(0.965\textwidth,1.7cm){\small 0.5}
  \end{picture}
  \caption{Normal orientation results of different methods. We show the point clouds colored by their normals and by per-point angular error values (Eq.~\eqref{eq:angular-error-pcd} without averaging across points). We also show the histograms of angular errors ($y$-axis in log-scale). $^\dag$Downsampled to 50,000 points. $^\ddag$Downsampled to 5,000 points.\label{fig:eval_normal}}
  \Description{Qualitative results of normal errors.}
\end{figure*}
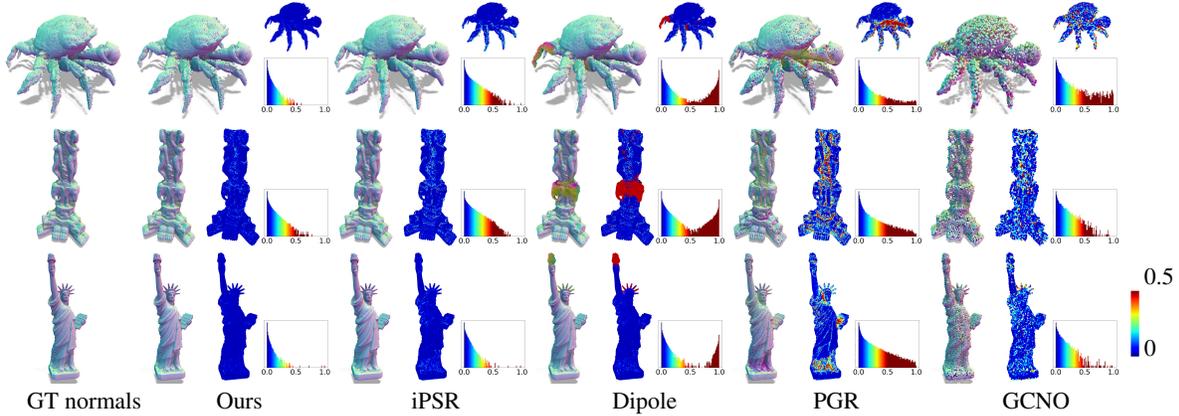

\subsubsection{Datasets}

Following previous work, we use datasets that are commonly used in surface reconstruction and normal orientation, including ABC~\citep{koch2019abc}, Thingi10k~\citep{Zhou2016Thingi10K}, the Stanford scanning repository~\citep{curless1996vmbcmri,krishnamurthy1996,turk1994,gardner2003}, and other datasets from \citet{lu2018gr,huang2019vipss,huang2022benchmark,erler2020p2s,metzer2021dipole,lin2023pgr,xu2023gcno}.

\begin{figure*}
  \centering
  \includegraphics[width=\textwidth]{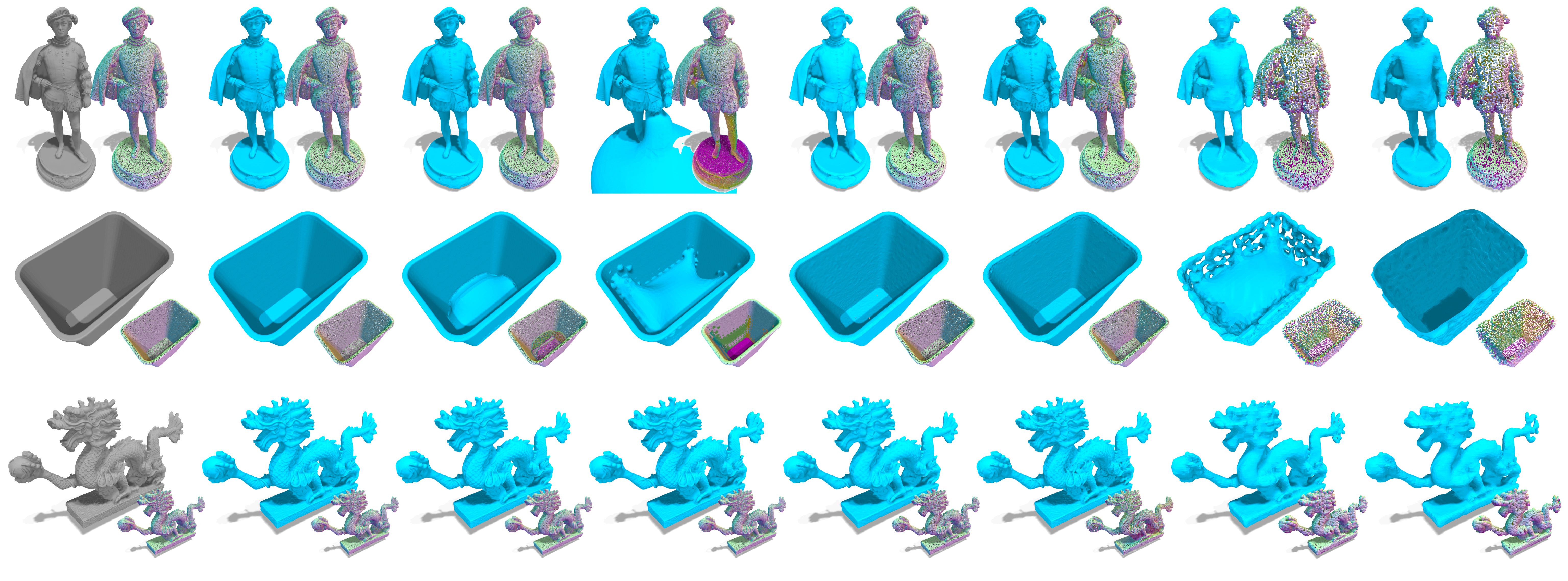}
  \begin{picture}(\textwidth,0.5cm)
  \put(0.060\textwidth,0.2cm){\small GT}
  \put(0.175\textwidth,0.2cm){\small Ours}
  \put(0.300\textwidth,0.2cm){\small iPSR}
  \put(0.425\textwidth,0.2cm){\small Dipole}
  \put(0.546\textwidth,0.2cm){\small Ours$^\dag$}
  \put(0.680\textwidth,0.2cm){\small PGR$^\dag$}
  \put(0.795\textwidth,0.2cm){\small Ours$^\ddag$}
  \put(0.917\textwidth,0.2cm){\small GCNO$^\ddag$}
  \end{picture}
  \caption{Normal orientation and surface reconstruction results of different methods. We show reconstructed meshes in plain colors and point clouds colored by their normals. $^\dag$Downsampled to 50,000 points. $^\ddag$Downsampled to 5,000 points.\label{fig:eval_mesh}}
  \Description{Qualitative results of normal orientation and surface reconstructions.}
\end{figure*}
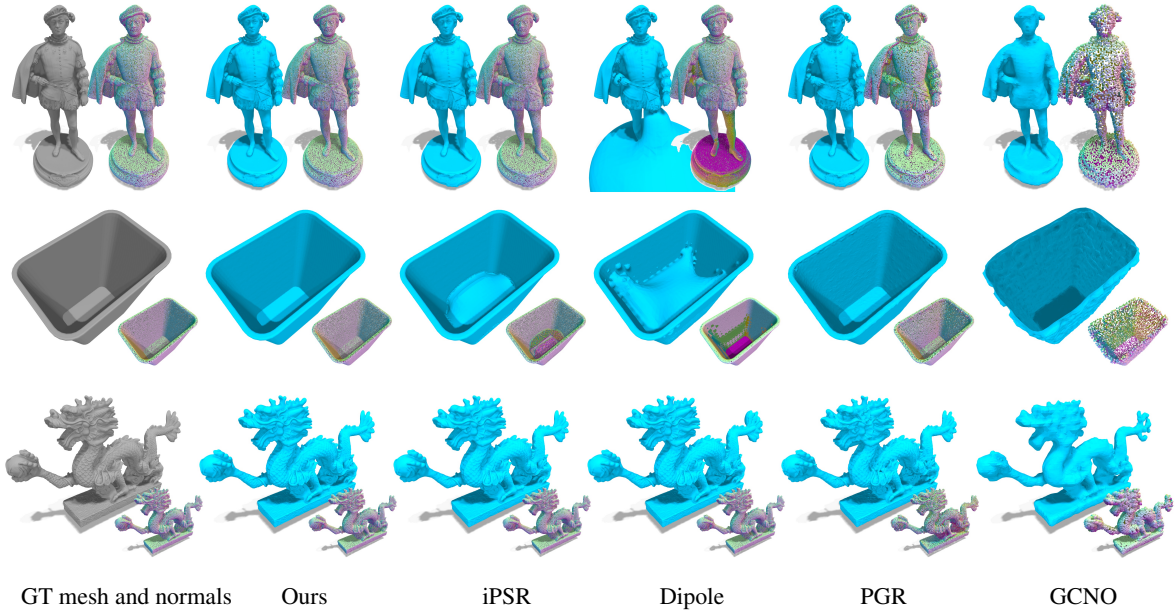

\begin{table*}
    \centering
    \caption{Quantitative comparison results of different methods evaluated with uniformly sampled models  exhibited in Fig.~\ref{fig:eval_normal} and Fig.~\ref{fig:eval_mesh}. The best results are marked in bold. $^\dag$Downsampled to 50,000 points. $^\ddag$Downsampled to 5,000 points.\label{tbl:eval_vis2}}
    \scriptsize
    \begin{tabular}{c|c|c|c|c|c|c|c|c|c|c|c|c|c|c|c}
        \toprule
        \multirow{2}{*}{Method} & \multirow{2}{*}{Model} & CD & ${\rm AE}_{\rm mesh}$ & ${\rm AE}_{\rm pcd}$ & $P_{\rm co}$ & \multirow{2}{*}{Model} & CD & ${\rm AE}_{\rm mesh}$ & ${\rm AE}_{\rm pcd}$ & $P_{\rm co}$ & \multirow{2}{*}{Model} & CD & ${\rm AE}_{\rm mesh}$ & ${\rm AE}_{\rm pcd}$ & $P_{\rm co}$\\
         & & ($\times10^{-4}$) & ($\times10^{-2}$) & ($\times10^{-2}$) & (\%) & & ($\times10^{-4}$) & ($\times10^{-2}$) & ($\times10^{-2}$) & (\%) & & ($\times10^{-4}$) & ($\times10^{-2}$) & ($\times10^{-2}$) & (\%) \\
        \midrule
        Dipole & & 15.1305 & 12.7237 & 12.2426 & 88.6625 & & 14.9317 & 13.5039 & 15.6297 & 86.2206 & & 4.8793 & 4.0816 & 4.6036 & 96.2181\\
        iPSR & & 2.0727 & 1.6770 & 1.9529 & 99.9356 & & 2.7776 & 2.8389 & 3.4912 & 99.8594 & & 1.5647 & 1.2212 & 1.3644 & 99.9963\\
        Ours & & \textbf{1.4556} & \textbf{1.3515} & \textbf{0.8401} & \textbf{99.9981} & \multirow{2}{*}{XYZRGB} & \textbf{2.0174} & \textbf{2.4180} & \textbf{1.7715} & \textbf{99.9781} & & \textbf{1.0802} & \textbf{1.0209} & \textbf{0.4961} & \textbf{99.9975}\\
        \cline{1-1}
        \cline{3-6}
        \cline{8-11}
        \cline{13-16}
        PGR$^\dag$ & Crab & 13.1959 & 11.5137 & 18.0785 & 86.8780 & \multirow{2}{*}{Statuette} & 13.5690 & 12.9512 & 21.8533 & 84.3820 & Liberty & 4.6810 & 4.3906 & 13.5943 & 93.2160 \\
        Ours$^\dag$ & & \textbf{3.4914} & \textbf{2.6714} & \textbf{1.8335} & \textbf{99.9300} & & \textbf{5.0047} & \textbf{4.8284} & \textbf{3.4662} & \textbf{99.8560} & & \textbf{3.4957} & \textbf{2.6623} & \textbf{1.2681} & \textbf{99.9880}\\
        \cline{1-1}
        \cline{3-6}
        \cline{8-11}
        \cline{13-16}
        GCNO$^\ddag$ & & 38.1440 & 11.3096 & 17.7151 & 88.4400 & & 29.7463 & 12.5233 & 9.3252 & \textbf{98.3400} & & 15.2068 & 7.2695 & 7.4188 & 99.2200\\
        Ours$^\ddag$ & & \textbf{20.7035} & \textbf{7.3027} & \textbf{5.7689} & \textbf{98.7800} & & \textbf{26.5084} & \textbf{12.2513} & \textbf{9.0382} & 98.0600 & & \textbf{12.5331} & \textbf{6.8335} & \textbf{5.9479} & \textbf{99.6800} \\
        \midrule
        Dipole & & 145.0533 & 30.4131 & 39.6944 & 60.5975 & & 34.4190 & 25.5114 & 22.1820 & 77.9050 & & 4.0927 & 5.7576 & 5.3654 & 95.5606\\
        iPSR & & 1.5918 & 1.0918 & 1.0650 & 99.9963 & & 23.9465 & 14.9586 & 13.9820 & 86.4475 & & 1.9667 & 5.1030 & 5.0781 & \textbf{96.3306}\\
        Ours & \multirow{2}{*}{Statue of} & \textbf{1.1420} & \textbf{0.9144} & \textbf{0.4398} & \textbf{99.9994} & \multirow{2}{*}{Trash} & \textbf{1.2967} & \textbf{0.3240} & \textbf{0.1370} & \textbf{100.0000} & & \textbf{1.3937} & \textbf{4.9502} & \textbf{4.1447} & \textbf{96.3306}\\
        \cline{1-1}
        \cline{3-6}
        \cline{8-11}
        \cline{13-16}
        PGR$^\dag$ & \multirow{2}{*}{Delaunay} & 7.4997 & 5.9834 & 11.5123 & 93.4080 & \multirow{2}{*}{Can} & 6.4437 & \textbf{0.6648} & 2.1229 & 99.6740 & Dragon & 7.9294 & 8.9578 & 19.2768 & 88.4840 \\
        Ours$^\dag$ & & \textbf{3.3201} & \textbf{2.1595} & \textbf{1.0475} & \textbf{99.9920} & & \textbf{6.0636} & 0.7209 & \textbf{0.3859} & \textbf{100.0000} & & \textbf{4.3697} & \textbf{5.7403} & \textbf{5.0017} & \textbf{96.3240}\\
        \cline{1-1}
        \cline{3-6}
        \cline{8-11}
        \cline{13-16}
        GCNO$^\ddag$ & & 24.6961 & 6.7809 & 5.2512 & 99.2000 & & \textbf{65.8343} & 44.8330 & 47.0361 & 53.9437 & & 18.3662 & 8.8840 & 8.7805 & \textbf{97.6000}\\
        Ours$^\ddag$ & & \textbf{20.6545} & \textbf{6.3045} & \textbf{4.2547} & \textbf{99.4400} & & 102.9334 & \textbf{36.8877} & \textbf{32.5923} & \textbf{70.7071} & & \textbf{15.2683} & \textbf{8.2227} & \textbf{7.5333} & 98.0800 \\
        \bottomrule
    \end{tabular}
\end{table*}

\subsubsection{Metrics}

We measure the quality of reconstructed normals with the following commonly used metrics. For point clouds with ground-truth normals $\{n_i^{\rm gt}\}_{i=1}^N$, we directly measure the angular error (AE) between reconstructed normals $\{n_i^{\rm recon}\}_{i=1}^N$ and ground-truth normals $\{n_i^{\rm gt}\}_{i=1}^N$:
\begin{equation}
    {\rm AE}_{\rm pcd}=\frac{1}{N}\sum_{i=1}^N(1-n_i^{\rm gt}\cdot n_i^{\rm recon})/2.\label{eq:angular-error-pcd}
\end{equation}
We also follow \citet{xu2023gcno} to use the percentage of correctly oriented normals as a metric:
\begin{equation}
    P_{\rm co}=\#\{i:n_i^{\rm recon}\cdot n_i^{\rm gt} > 0\}/N.
\end{equation}
Another common practice is to measure the accuracy of normals using their reconstructed meshes. For fairness, we use Screened Poisson Surface Reconstruction (SPSR)~\citep{kazhdan2013spr} for all methods. We set the maximum tree depth in SPSR to 10 and keep other parameters as default. We use the mesh Chamfer distance (CD) and mesh normal angular error (${\rm AE}_{\rm mesh}$). Here, ${\rm CD}$ for two meshes $M_1, M_2$ is defined as
\begin{equation}
    {\rm CD}=\frac{1}{2}\Big(\frac{1}{N}\sum_{i=1}^N\|x_i-\hat x_i\|+\frac{1}{N}\sum_{i=1}^N\|y_i-\hat y_i\|\Big),
\end{equation}
where $\{x_i\}_{i=1}^N$ uniformly samples $M_1$ and $\hat x_i$ is its closest point on $M_2$ (same for $y_i\in M_2$ and $\hat y_i\in M_1$). The expression for ${\rm AE}_{\rm mesh}$ is the same as ${\rm AE}_{\rm pcd}$, except it is a two-way average between the normals of the above closest samples on the meshes.

\subsection{Evaluation of Normal Quality}

\subsubsection{Accuracy}

We use uniform samples without noise to measure the accuracy of different methods. For each mesh, we use \verb|sample_surface_even| in Trimesh~\citep{trimesh} to obtain uniform samples. Each model is sampled with 160,000 points. Note that PGR~\citep{lin2023pgr} and GCNO~\citep{xu2023gcno} have prohibitively high memory and time consumption, respectively. Hence, we downsample the input point cloud to 50,000 points for PGR and 5,000 points for GCNO. Downsampled experiments are denoted by X$^\dag$ (50,000 points) and X$^\ddag$ (5,000 points), where X is the method name.

\begin{table}
    \centering
    \caption{Quantitative comparison results of different methods on a large-scale benchmark dataset~\citep{huang2022benchmark}. Each metric is an average over all 1387 models used for this evaluation. The best results are in bold. $^\dag$Downsampled to 50,000 points.\label{tbl:eval_benchmark}}
    \begin{tabular}{c|c|c|c|c}
        \toprule
        \multirow{2}{*}{Method} & CD & ${\rm AE}_{\rm mesh}$ & ${\rm AE}_{\rm pcd}$ & $P_{\rm co}$ \\
         & ($\times10^{-4}$) & ($\times10^{-2}$) & ($\times10^{-2}$) & (\%) \\
        \midrule
        Dipole & 7.7909 & 5.3215 & 5.0029 & 95.2759 \\
        iPSR & 1.9306 & 0.9215 & 0.6108 & 99.8208 \\
        Ours & \textbf{1.1274} & \textbf{0.8104} & \textbf{0.3097} & \textbf{99.9328} \\
        \midrule
        PGR$^\dag$ & 2.6093 & 1.3621 & 3.1033 & 99.3609 \\
        Ours$^\dag$ & \textbf{2.5337} & \textbf{1.0878} & \textbf{0.5418} & \textbf{99.8683} \\
        \bottomrule
    \end{tabular}
\end{table}

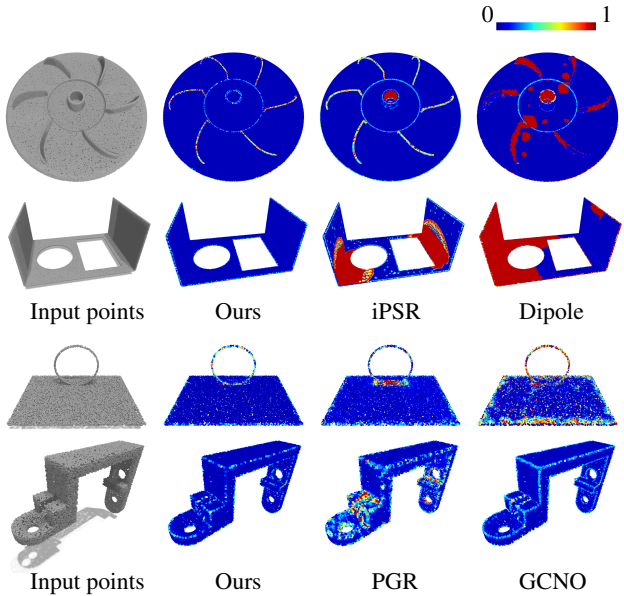
\begin{figure}
  \centering
  \includegraphics[width=\linewidth]{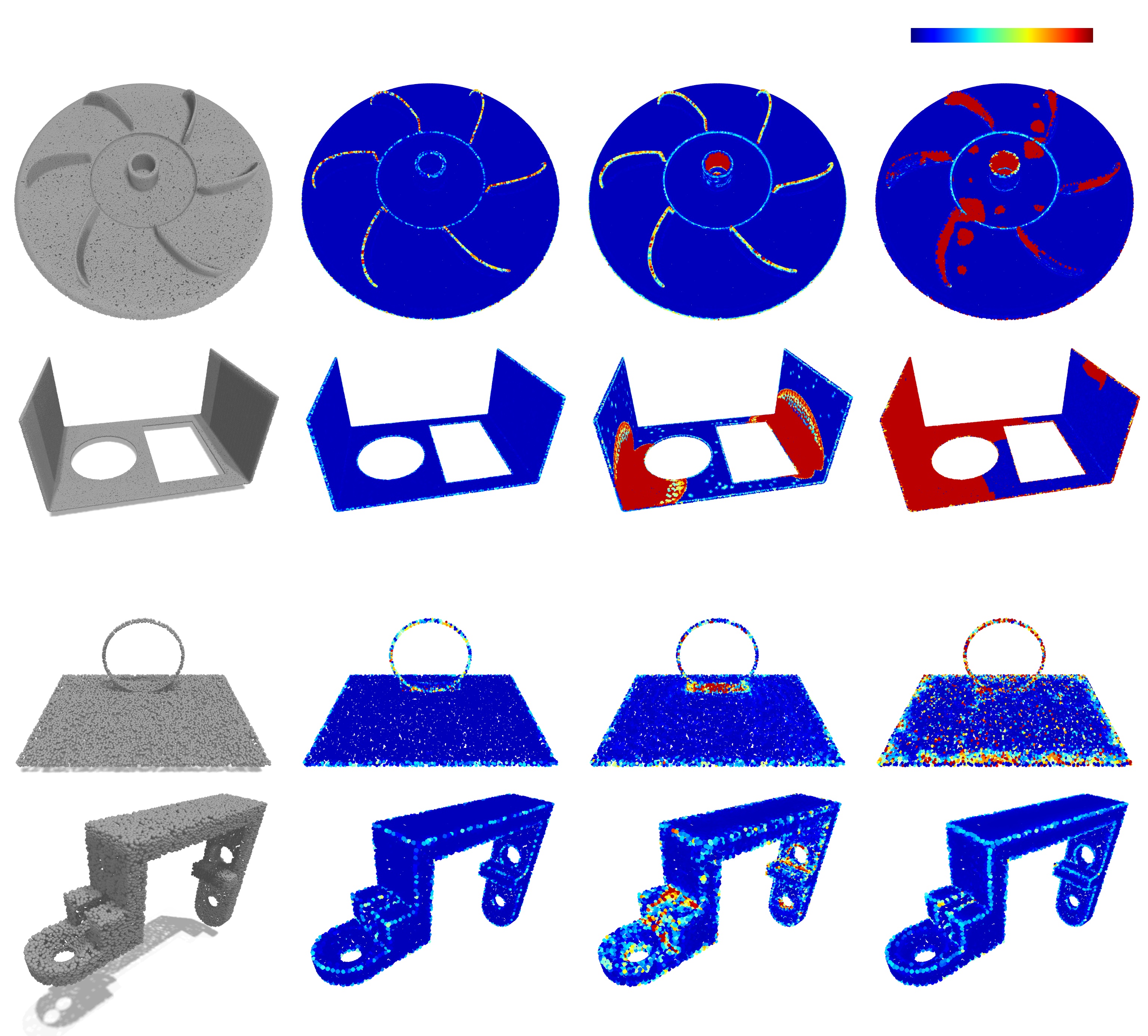}
  \begin{picture}(\linewidth,0.3cm)
  \put(0.05\linewidth,3.8cm){\small Input points}
  \put(0.345\linewidth,3.8cm){\small Ours}
  \put(0.595\linewidth,3.8cm){\small iPSR}
  \put(0.83\linewidth,3.8cm){\small Dipole}

  \put(0.05\linewidth,0.2cm){\small Input points}
  \put(0.345\linewidth,0.2cm){\small Ours}
  \put(0.595\linewidth,0.2cm){\small PGR}
  \put(0.83\linewidth,0.2cm){\small GCNO}
  
  \put(0.77\linewidth,7.75cm){\small 0}
  \put(0.96\linewidth,7.75cm){\small 1}
  \end{picture}
  \caption{Qualitative comparisons for models with thin structures and sharp edges. Each model in the first two rows has 100,000 points, and each in the last two rows has 10,000 points.\label{fig:thin_sharp}}
\end{figure}

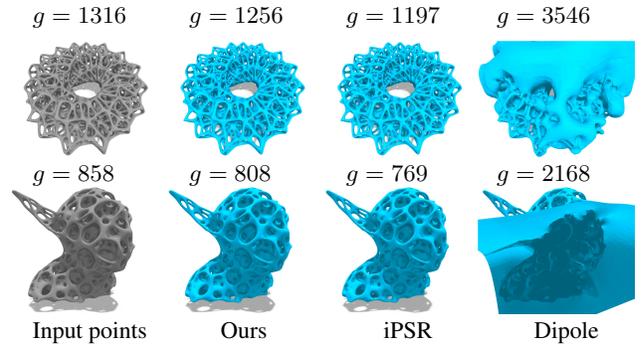
\begin{figure}
  \centering
  \includegraphics[width=\linewidth]{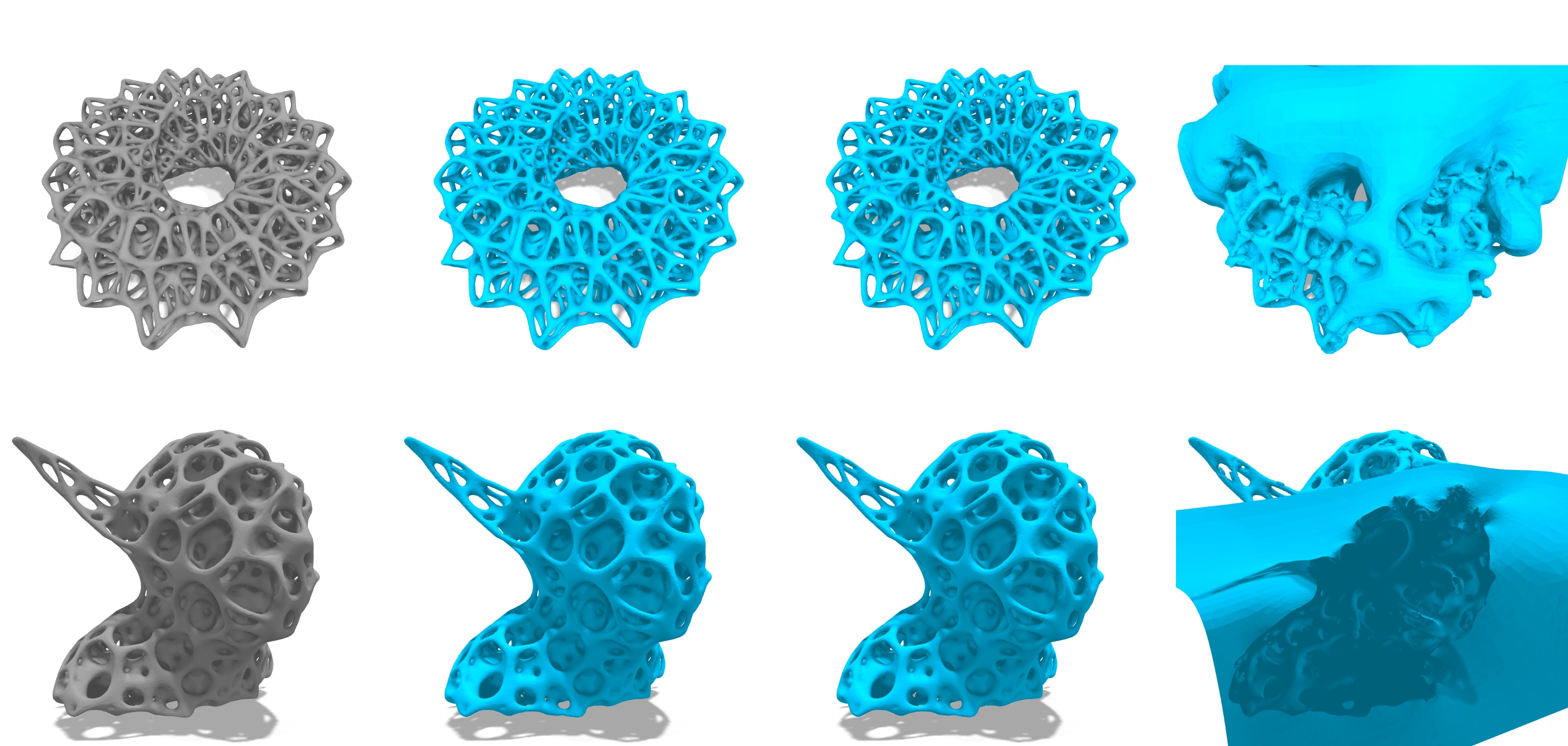}
  \begin{picture}(\linewidth,0.5cm)
  \put(0.04\linewidth,0.2cm){\small Input points}
  \put(0.34\linewidth,0.2cm){\small Ours}
  \put(0.60\linewidth,0.2cm){\small iPSR}
  \put(0.84\linewidth,0.2cm){\small Dipole}

  \put(0.04\linewidth,4.4cm){\small $g=1316$}
  \put(0.29\linewidth,4.4cm){\small $g=1256$}
  \put(0.54\linewidth,4.4cm){\small $g=1197$}
  \put(0.78\linewidth,4.4cm){\small $g=3546$}

  \put(0.04\linewidth,2.3cm){\small $g=858$}
  \put(0.29\linewidth,2.3cm){\small $g=808$}
  \put(0.54\linewidth,2.3cm){\small $g=769$}
  \put(0.78\linewidth,2.3cm){\small $g=2168$}

  \end{picture}
  \caption{Reconstruction results of two high-genus surfaces. We additionally show the genus $g$ of the GT models and the reconstructed meshes.\label{fig:highgenus}}
  \Description{Reconstruction results of high-genus surfaces.}
\end{figure}

\begin{figure*}
  \centering
  \includegraphics[width=\linewidth]{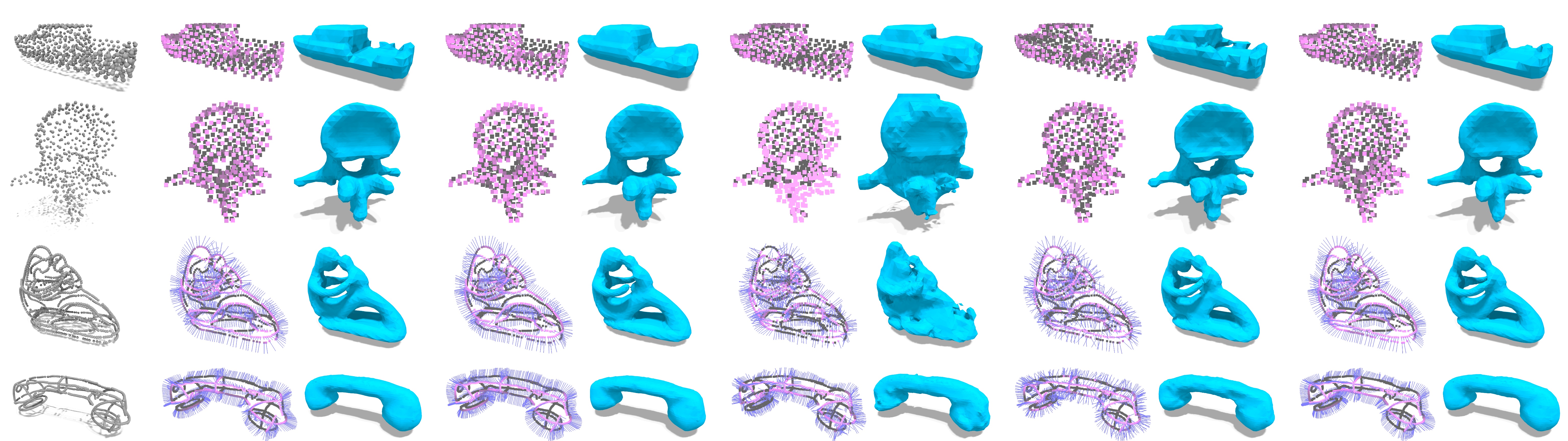}
  \begin{picture}(\linewidth,0.5cm)
  \put(0.0\linewidth,0.2cm){\small Input points}
  \put(0.175\linewidth,0.2cm){\small Ours}
  \put(0.360\linewidth,0.2cm){\small iPSR}
  \put(0.535\linewidth,0.2cm){\small Dipole}
  \put(0.725\linewidth,0.2cm){\small PGR}
  \put(0.895\linewidth,0.2cm){\small GCNO}
  \end{picture}
  \caption{Normal estimation and reconstruction results of sparse point clouds and 3D sketches from \citet{huang2019vipss}. Oriented points are rendered as point sprites, and we additionally render normals for 3D sketches.\label{fig:sparse_sketch}}
  \Description{Normal estimation and reconstruction results of sparse point clouds and 3D sketches.}
\end{figure*}

Fig.~\ref{fig:eval_normal} shows the oriented point clouds, their angular error maps and angular error distribution. It can be seen that dipole propagation~\citep{metzer2021dipole} has a high probability of incorrectly orienting normals in a large area. For other methods, while they can generally achieve a globally consistent orientation, their estimated normals have larger errors compared to our results. Fig.~\ref{fig:eval_mesh} shows reconstructed meshes using normals estimated by different methods. Note that almost all baselines cannot handle convex thin structures (see the interior of the trash can model). The quantitative results for these six models are reported in Table~\ref{tbl:eval_vis2}. Our method achieves the best overall performance. Note that for the downsampled Trash Can model with 5,000 points, the low sampling rate is insufficient to depict the underlying thin-layer geometry, and thus neither our method nor GCNO could work.

To further validate the accuracy of our estimated normals, we conduct a large-scale study on a new surface reconstruction benchmark dataset~\citep{huang2022benchmark}. Since we found the point clouds provided by the \citet{huang2022benchmark} contain noise (in both point positions and normals) and are unfit for quantitative evaluation, we use their provided meshes and sample uniform points using Trimesh~\citep{trimesh}. Furthermore, some meshes provided by this dataset are non-manifold with self-intersecting components, which are not suitable for either our method or the baselines. Hence, we remove all non-manifold data and use the remaining 1387 models for this large-scale evaluation. Each model is sampled with 160,000 points. Note that PGR~\citep{lin2023pgr} is memory-intensive, and thus we use inputs downsampled to 50,000 points for this evaluation. Since GCNO~\citep{xu2023gcno} is too computationally expensive (it would take months even with downsampled points), it is excluded from this comparison. Table~\ref{tbl:eval_benchmark} reports the metrics averaged across all 1387 models. Our method outperforms the baselines in a statistical sense.

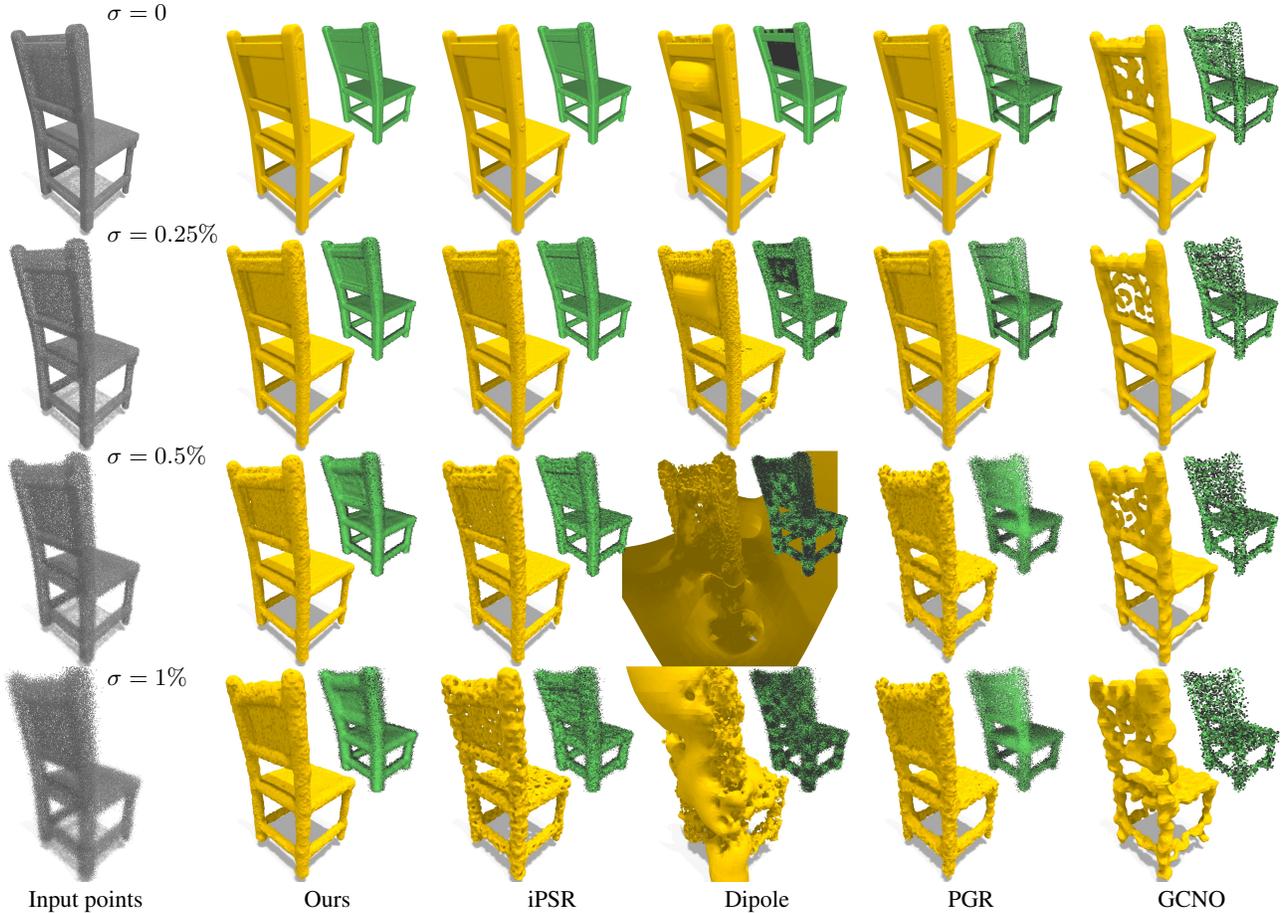
\begin{figure*}
  \centering
  \includegraphics[width=\textwidth]{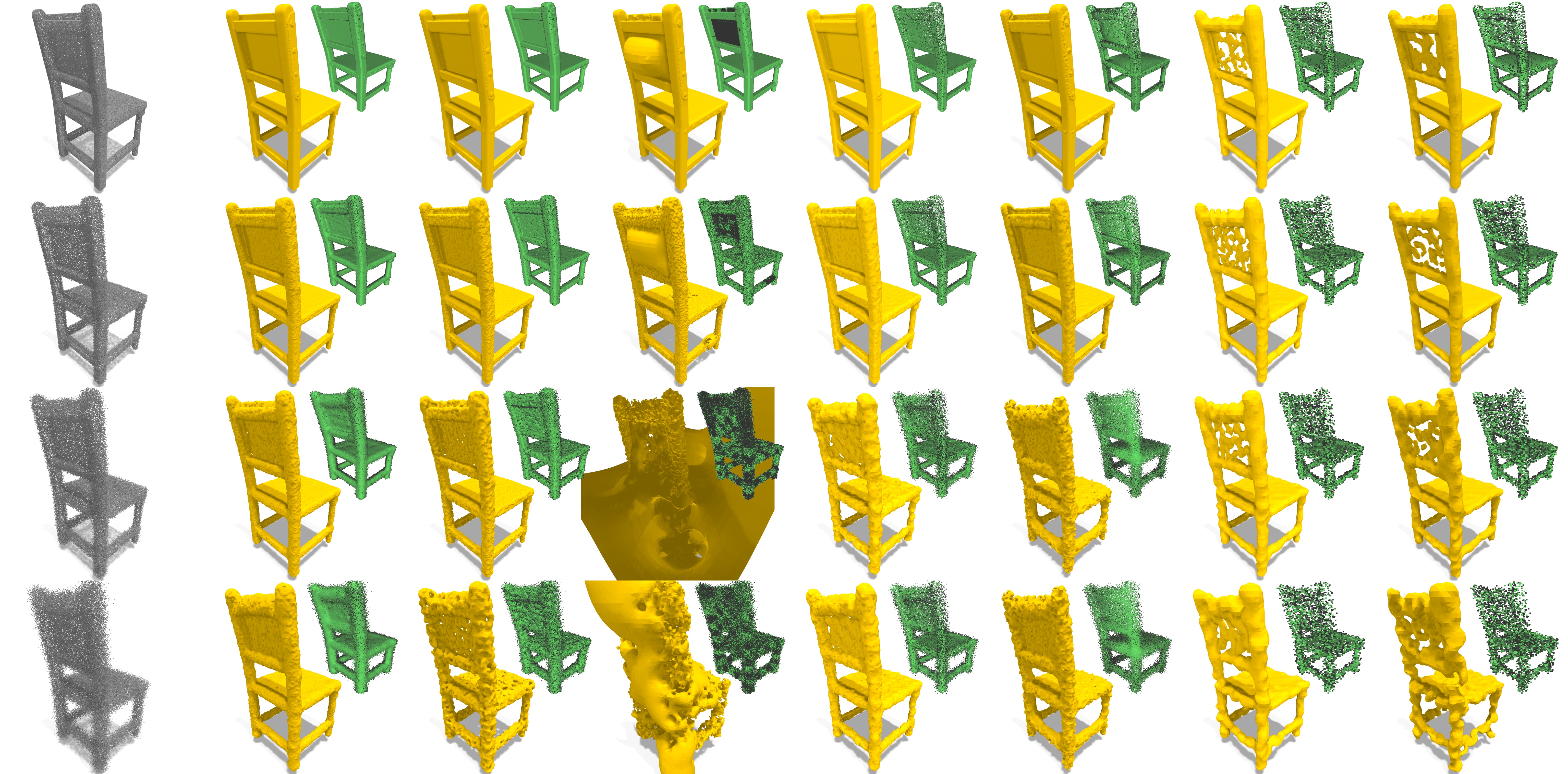}
  \begin{picture}(\textwidth,0.5cm)
  \put(0.04\textwidth,0.2cm){\small Input points}
  \put(0.19\textwidth,0.2cm){\small Ours}
  \put(0.31\textwidth,0.2cm){\small iPSR}
  \put(0.43\textwidth,0.2cm){\small Dipole}
  \put(0.56\textwidth,0.2cm){\small Ours$^\dag$}
  \put(0.69\textwidth,0.2cm){\small PGR$^\dag$}
  \put(0.81\textwidth,0.2cm){\small Ours$^\ddag$}
  \put(0.93\textwidth,0.2cm){\small GCNO$^\ddag$}

  \put(0.08\textwidth,8.6cm){\small $\sigma=0$}
  \put(0.08\textwidth,6.4cm){\small $\sigma=0.25\%$}
  \put(0.08\textwidth,4.2cm){\small $\sigma=0.5\%$}
  \put(0.08\textwidth,2cm){\small $\sigma=1\%$}
  \end{picture}
  \caption{Reconstruction (in yellow) and normal estimation results (in green) for a chair model with different noise levels. The normals are rendered as point sprites to indicate their orientation. $^\dag$Downsampled to 50,000 points. $^\ddag$Downsampled to 5,000 points.\label{fig:noise}}
  \Description{Reconstruction and normal estimation results for a chair model with different noise levels}
\end{figure*}

\begin{figure*}
  \centering
  \includegraphics[width=\textwidth]{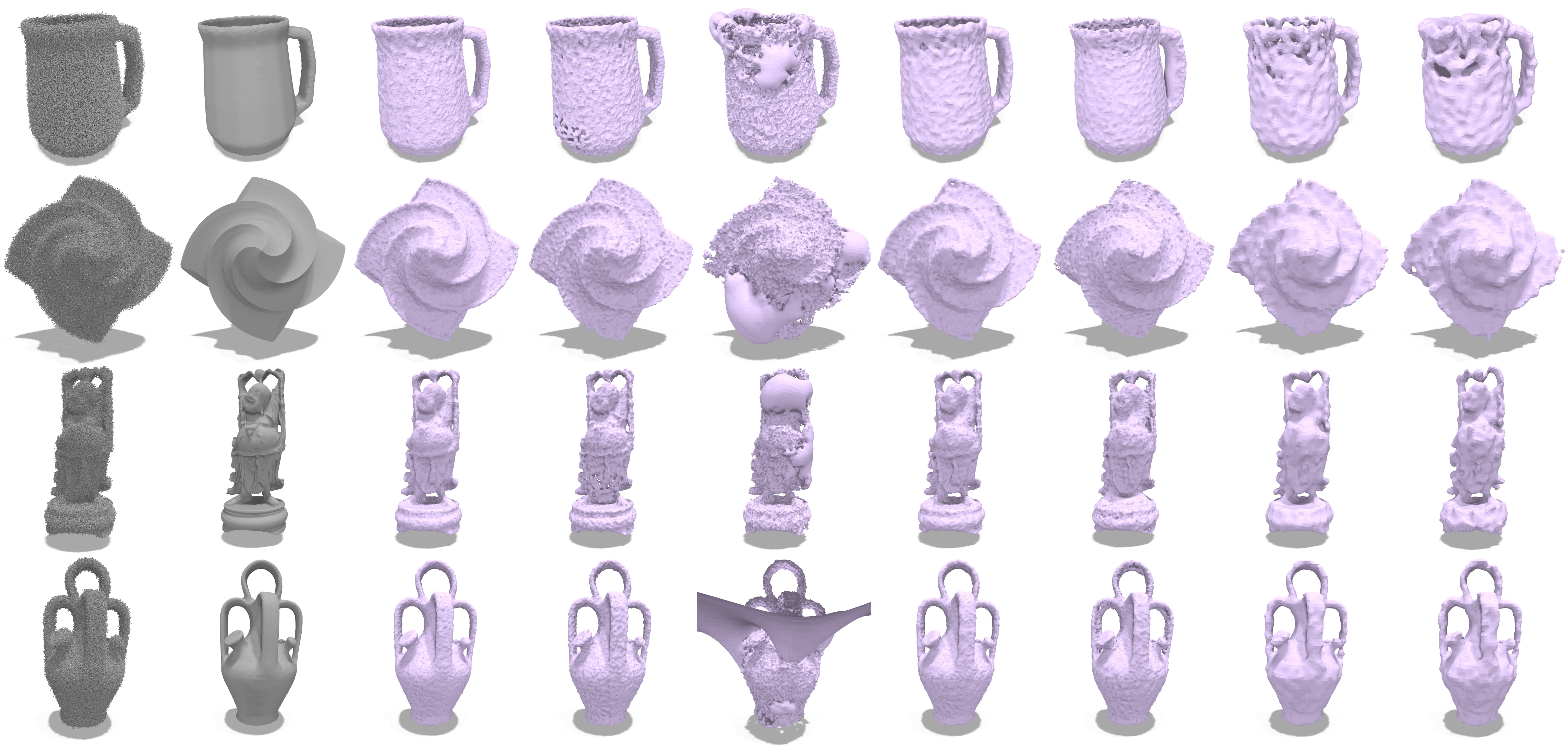}
  \begin{picture}(\textwidth,0.5cm)
  \put(0.01\textwidth,0.2cm){\small Input points}
  \put(0.14\textwidth,0.2cm){\small GT mesh}
  \put(0.26\textwidth,0.2cm){\small Ours}
  \put(0.37\textwidth,0.2cm){\small iPSR}
  \put(0.475\textwidth,0.2cm){\small Dipole}
  \put(0.592\textwidth,0.2cm){\small Ours$^\dag$}
  \put(0.705\textwidth,0.2cm){\small PGR$^\dag$}
  \put(0.815\textwidth,0.2cm){\small Ours$^\ddag$}
  \put(0.922\textwidth,0.2cm){\small GCNO$^\ddag$}
  \end{picture}
  \caption{Reconstruction results for four noisy point clouds. Each model is obtained by adding $\sigma=0.5\%$ Gaussian noise to uniform samples. $^\dag$Downsampled to 50,000 points. $^\ddag$Downsampled to 5,000 points.\label{fig:noise2}}
  \Description{}
\end{figure*}

\subsubsection{Thin Structures and Sharp Edges}

Thin structures and sharp edges constitute a difficult challenge for normal estimation methods, since normals tend to have abrupt changes around these structures.  These structures also commonly exist in real-world objects, e.g., CAD models. Fig.~\ref{fig:thin_sharp} shows the angular errors for four models with thin structures or sharp edges. Our method can produce correct normals even for very thin plates, and has a lower error near sharp edges compared to other methods.

\subsubsection{High-genus Surfaces}

We experiment with two high-genus surfaces shown in Fig.~\ref{fig:highgenus}. Due to their large surface areas, we sample 200,000 for each model to ensure the point clouds can depict the underlying geometric structure. We empirically found that our method and iPSR require more iterations to converge, and thus set the number of iterations for both methods to 300. In Fig.~\ref{fig:highgenus} we show the reconstruction results as well as the genus of each mesh. Our method can more faithfully reconstruct topological structures than other methods.

\subsubsection{Sparse Points and 3D Sketches}

Sparse points can be another challenge due to low sampling rates. Additionally, \citet{huang2019vipss} studied an extreme case of sparse inputs: 3D sketches, where the points are not only sparse but also highly non-uniform. In Fig.~\ref{fig:sparse_sketch} we show four models from \citet{huang2019vipss}. The two uniform sparse point clouds each have 500 points, and the two sketches each have 1,000 points. Our method can obtain plausible normals for both types of point clouds.

\begin{table*}
    \centering
    \caption{Quantitative comparisons of estimating normals for noisy models exhibited in Fig.~\ref{fig:noise} and Fig.~\ref{fig:noise2}. Since it does not make sense to talk about ``GT normals'' of noisy points, we measure only mesh errors. The best results are marked in bold. $^\dag$Downsampled to 50,000 points. $^\ddag$Downsampled to 5,000 points.\label{tbl:eval_noise_both}}
    \small
    \begin{tabular}{c|c|c|c|c|c|c|c|c|c|c|c|c|c}
        \toprule
        & \multirow{2}{*}{Method} 
        & Noise & CD & ${\rm AE}_{\rm mesh}$  
        & Noise & CD & ${\rm AE}_{\rm mesh}$
        & Noise & CD & ${\rm AE}_{\rm mesh}$
        & Noise & CD & ${\rm AE}_{\rm mesh}$ \\
        & & level $\sigma$ & ($\times10^{-3}$) & ($\times10^{-1}$) 
        & level $\sigma$ & ($\times10^{-3}$) & ($\times10^{-1}$) 
        & level $\sigma$ & ($\times10^{-3}$) & ($\times10^{-1}$) 
        & level $\sigma$ & ($\times10^{-3}$) & ($\times10^{-1}$)  \\
        \midrule
        & Dipole & & 1.2309 & 0.5431 & &  2.0284 & 1.5253 & & 22.6848 & 5.1618 & &  14.1688 & 4.6009\\
        & iPSR & & 0.2434 & 0.1186 & & \textbf{0.8082} & 0.3545  & &  1.7020 & 0.6713 & &  6.3857 & 3.1309\\
        & Ours & & \textbf{0.1899} & \textbf{0.0957} & & 0.8242 & \textbf{0.3520} & & \textbf{1.6124} & \textbf{0.5665} & & \textbf{3.6541} & \textbf{0.8072}\\
        \cline{2-2}
        \cline{4-5}
        \cline{7-8}
        \cline{10-11}
        \cline{13-14}
        Fig.~\ref{fig:noise} & PGR$^\dag$ & $0$  & 1.5540 & 0.2830 & $0.25\%$ & 1.5270 & 0.5409  & $0.5\%$ &  9.3366 & 1.8209 & $1\%$ & 6.6418 & 1.5744\\
        & Ours$^\dag$ & & \textbf{0.3961} & \textbf{0.1694} & & \textbf{1.0456} & \textbf{0.3223}  & &  \textbf{4.9580} & \textbf{0.1694} & &  \textbf{3.4486} & \textbf{0.3223}\\
        \cline{2-2}
        \cline{4-5}
        \cline{7-8}
        \cline{10-11}
        \cline{13-14}
        & GCNO$^\ddag$ & & 4.1716 & 0.5879 & & 2.8376 & 0.6831  & &  3.9593 & 0.9416 & &  7.7393 & 1.8608\\
        & Ours$^\ddag$ & & \textbf{2.5960} & \textbf{0.5502} & & \textbf{2.5552} & \textbf{0.6374}  & &  \textbf{3.3851} & \textbf{0.5502} & &  \textbf{5.3490} & \textbf{0.6374}\\
        \midrule
        & \multirow{2}{*}{Method} 
        & \multirow{2}{*}{Model} & CD & ${\rm AE}_{\rm mesh}$  
        & \multirow{2}{*}{Model} & CD & ${\rm AE}_{\rm mesh}$ 
        & \multirow{2}{*}{Model} & CD & ${\rm AE}_{\rm mesh}$ 
        & \multirow{2}{*}{Model} & CD & ${\rm AE}_{\rm mesh}$  \\
        &  & & ($\times10^{-3}$) & ($\times10^{-2}$)  
         & & ($\times10^{-3}$) & ($\times10^{-2}$) 
         & & ($\times10^{-3}$) & ($\times10^{-2}$) 
         & & ($\times10^{-3}$) & ($\times10^{-2}$)  \\
        \midrule
        & Dipole & & 5.7139 & 23.4262 & & 6.4231 & 29.9646 & & 5.1984 & 32.2373 & & 20.2586 & 47.9005 \\
        & iPSR & & 3.5387 & 8.5899 & & \textbf{1.7867} & 4.9125  & &  2.4111 & 13.7234 & &  1.4456 & 3.7068\\
        & Ours & & \textbf{1.9513} & \textbf{2.9340} & & 1.8952 & \textbf{4.1823} & & \textbf{2.1011} & \textbf{8.8583} & & \textbf{1.0311} & \textbf{2.3479}  \\
        \cline{2-2}
        \cline{4-5}
        \cline{7-8}
        \cline{10-11}
        \cline{13-14}
        Fig.~\ref{fig:noise2} & PGR$^\dag$ & Cup & 7.3970 & 10.3166 & Flower & 3.2521 & 4.8484  & Buddha &  4.4460 & 18.8665 & Vase &  3.0096 & 8.6615\\
        & Ours$^\dag$ & & \textbf{2.5295} & \textbf{2.2244} & & \textbf{1.9771} & \textbf{3.8066}  & &  \textbf{2.3528} & \textbf{8.2716} & &  \textbf{1.3408} & \textbf{2.3336}\\
        \cline{2-2}
        \cline{4-5}
        \cline{7-8}
        \cline{10-11}
        \cline{13-14}
        & GCNO$^\dag$ & & 8.3960 & 9.8521 & & 3.7648 & 5.4055  & &  4.2233 & 11.8817 & &  3.0652 & 3.2314 \\
        & Ours$^\ddag$ & & \textbf{5.0091} & \textbf{5.2786} & & \textbf{3.3762} & \textbf{4.8993}  & &  \textbf{3.8669} & \textbf{10.8446} & &  \textbf{2.6241} & \textbf{2.6057}\\
        \bottomrule
    \end{tabular}
\end{table*}

\begin{figure*}
  \centering
  \includegraphics[width=\linewidth]{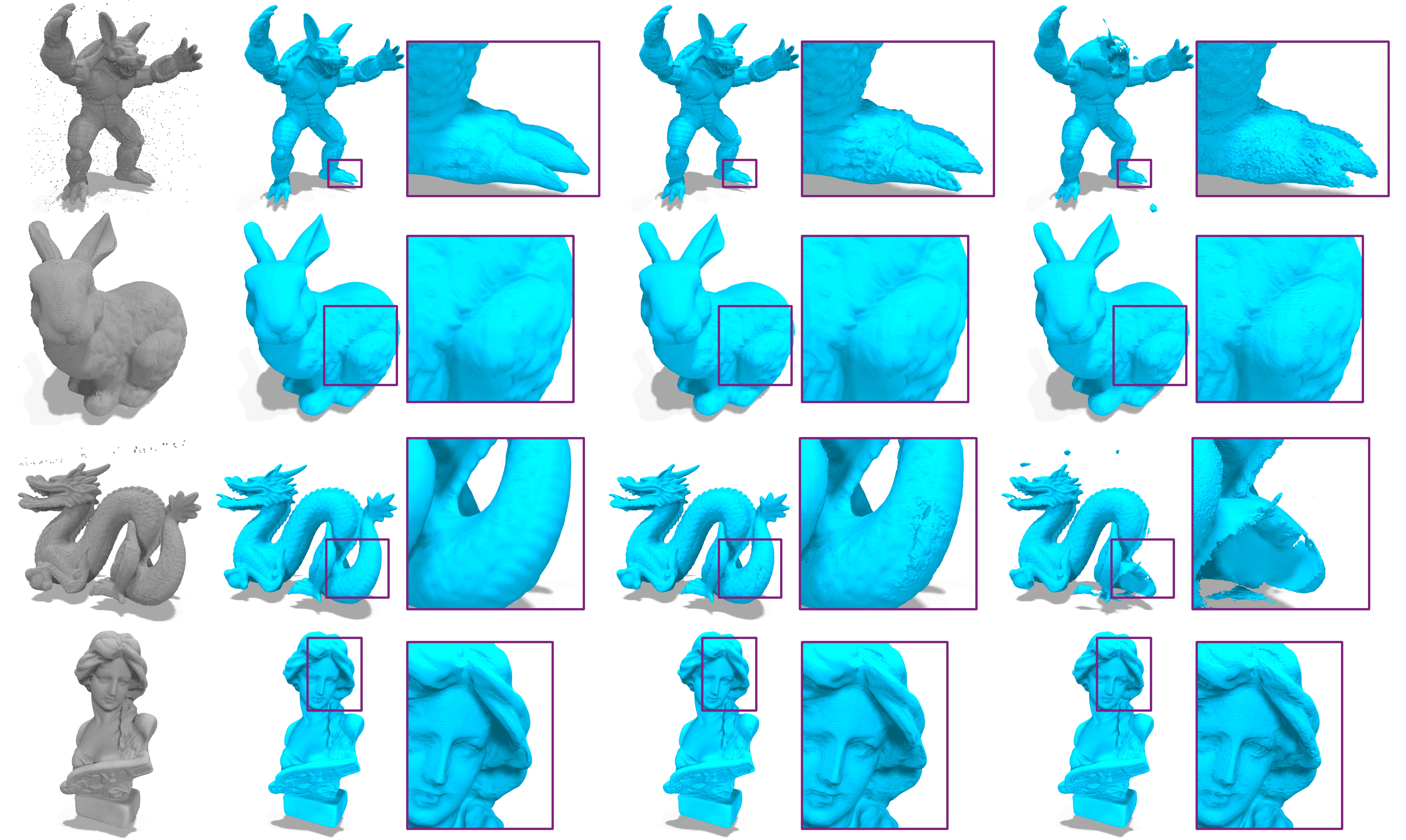}
  \begin{picture}(\linewidth,0.5cm)
  \put(0.04\linewidth,0.2cm){\small Input points}
  \put(0.27\linewidth,0.2cm){\small Ours}
  \put(0.555\linewidth,0.2cm){\small iPSR}
  \put(0.827\linewidth,0.2cm){\small Dipole}
  \end{picture}
  \caption{Reconstruction results of raw range scans from the Stanford scanning repository~\citep{curless1996vmbcmri,krishnamurthy1996,turk1994,gardner2003} and \citet{lu2018gr}. Number of points in each model: 2,374,290 (Armadillo); 362,272 (Bunny); 2,109,047 (Dragon); 530,497 (Lady). Please zoom in to see scanner noise patterns and reconstruction details.\label{fig:scans}}
  \Description{Reconstruction results of raw range scans.}
\end{figure*}

\begin{figure*}
  \centering
  \includegraphics[width=\textwidth]{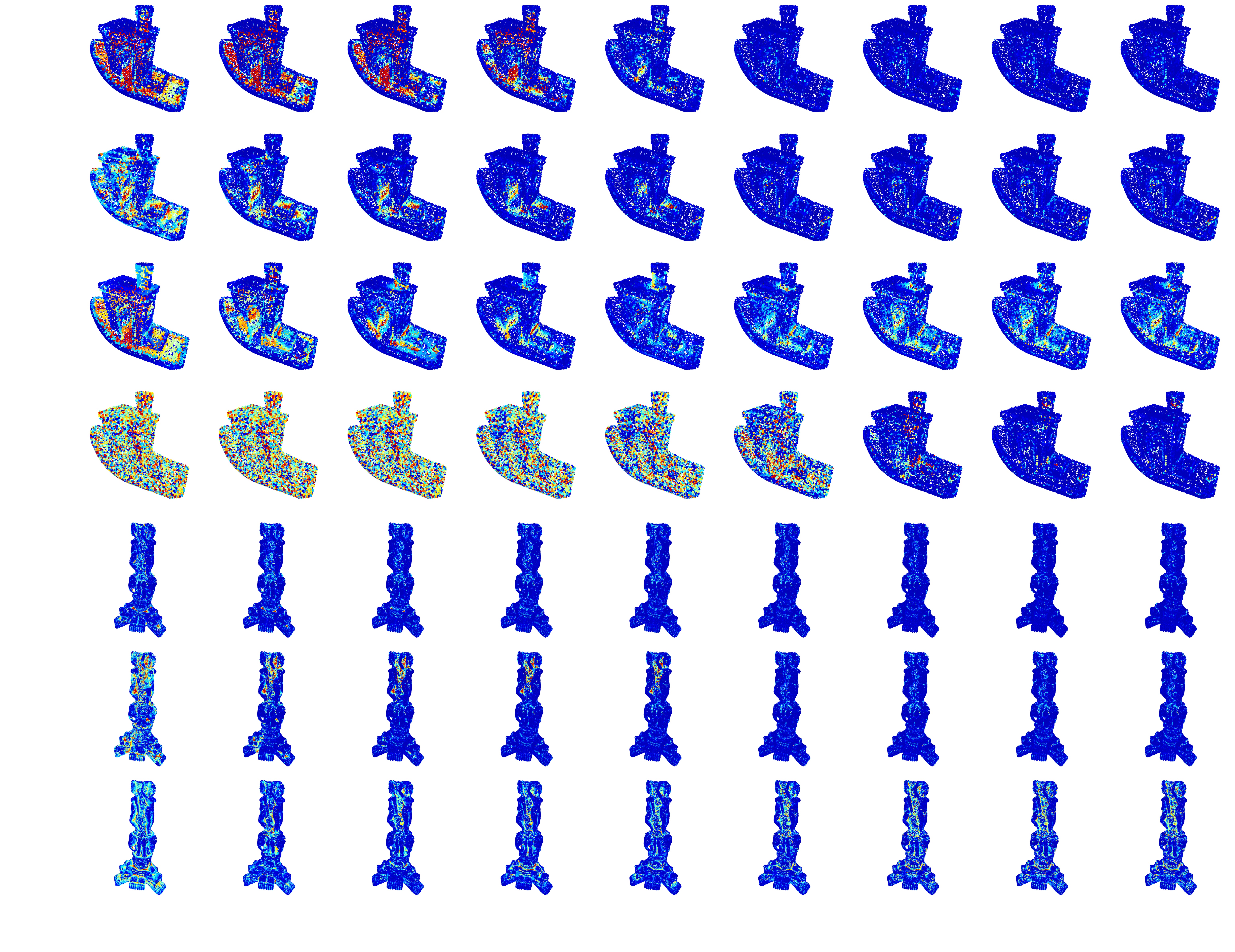}
  \begin{picture}(\textwidth,0.0cm)
  \put(0.02\textwidth,13.3cm){\small Ours}
  \put(0.02\textwidth,11.45cm){\small iPSR}
  \put(0.02\textwidth,9.6cm){\small PGR}
  \put(0.02\textwidth,7.75cm){\small GCNO}
  \put(0.02\textwidth,5.9cm){\small Ours}
  \put(0.02\textwidth,4.05cm){\small iPSR}
  \put(0.02\textwidth,2.1cm){\small PGR}
  
  \put(0.01\textwidth,0.8cm){\small \#Iterations}

  \put(0.11\textwidth,0.8cm){\small 1}
  \put(0.21375\textwidth,0.8cm){\small 2}
  \put(0.3175\textwidth,0.8cm){\small 3}
  \put(0.42125\textwidth,0.8cm){\small 4}
  \put(0.525\textwidth,0.8cm){\small 5}
  \put(0.62875\textwidth,0.8cm){\small 10}
  \put(0.7325\textwidth,0.8cm){\small 20}
  \put(0.83625\textwidth,0.8cm){\small 30}
  \put(0.94\textwidth,0.8cm){\small 40}

  \end{picture}
  \caption{Illustration of the convergence processes of different methods when applied to two uniformly sampled models (3DBenchy and XYZRGB Statuette). We visualize each intermediate result using error maps. Red indicates high error and blue indicates low error. We show the first 40 iterations of each method.\label{fig:itertest_all}}
  \Description{Error color maps of convergence process.}
\end{figure*}

\begin{figure*}
  \centering
  \includegraphics[width=\textwidth]{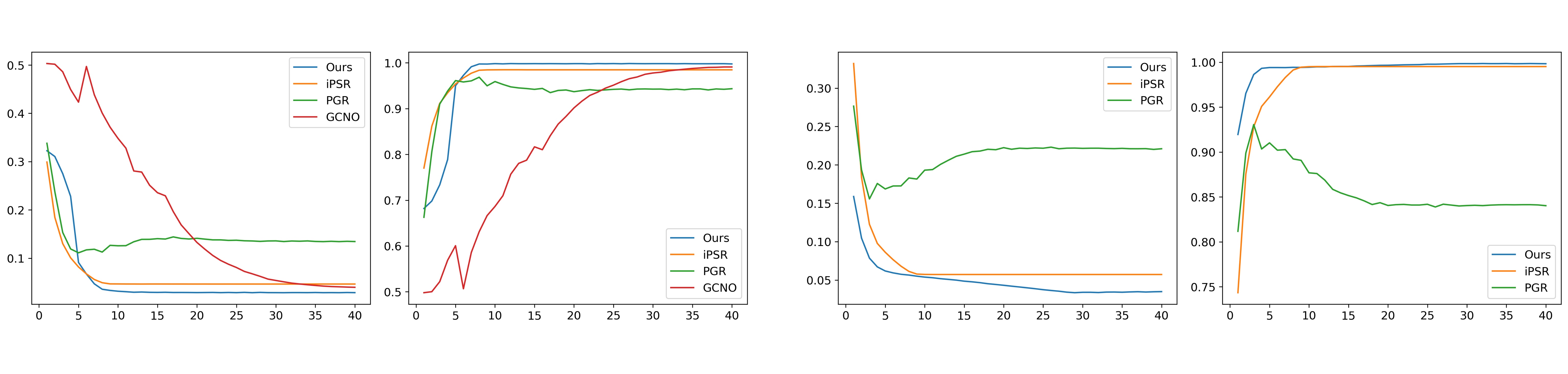}
  \begin{picture}(\textwidth,0.0cm)

  \put(0.09\textwidth,4.3cm){\small Angular Error}
  \put(0.36\textwidth,4.3cm){\small $P_{\rm co}$}

  \put(0.60\textwidth,4.3cm){\small Angular Error}
  \put(0.88\textwidth,4.3cm){\small $P_{\rm co}$}
  
  \put(0.21\textwidth,0.6cm){\small 3DBenchy}
  \put(0.7\textwidth,0.6cm){\small XYZRGB Statuette}

  \end{picture}
  \caption{Plots of the average angular error and the percentage of correctly oriented normals ($P_{\rm co}$) of the convergence processes of different methods.\label{fig:itertest_quanti}}
  \Description{Plots of the average angular error and the percentage of correctly oriented normals ($P_{\rm co}$) of the convergence processes of different method Illustration of the convergence process of different methods.}
\end{figure*}

\subsubsection{Noise Resilience}

We study the noise resilience of different methods by adding Gaussian noise to clean uniform samples. We use three noise levels $\sigma=0.25\%,0.5\%,1\%$ (w.r.t. the bounding box diagonal) as the standard deviation for Gaussian noise. Following previous settings, each model is sampled with 160,000 points, which are downsampled to 50,000 for PGR~\citep{lin2023pgr} and 5,000 for GCNO~\citep{xu2023gcno}. We noticed that default settings of several methods do not work well with noisy inputs, and therefore we adjust the parameters to obtain optimal results. For our method, we set $w_1=0.02,0.03,0.04$ and $w_2=4w_1$ for the three noise levels. For iPSR~\citep{hou2022ipsr}, we set the point weight to 0.5. For PGR~\citep{lin2023pgr}, we use their recommended settings for noisy data. Note that Dipole~\citep{metzer2021dipole} and GCNO~\citep{xu2023gcno} do not have noise-adapting options or parameters, and we keep their default settings. We also change the point weight to 0.5 for the final reconstruction step using SPSR~\citep{kazhdan2013spr}.

Fig.~\ref{fig:noise} shows the reconstructed meshes (in yellow) and the estimated normals (in green) for different noise levels. For dipole propagation~\citep{metzer2021dipole}, even without noise it cannot infer normals with globally consistent orientation, and as noise level increases its normals become completely misoriented. Other methods can infer a roughly correct orientation but are notably worse at preserving the geometric structure, especially at high noise levels. Fig.~\ref{fig:noise2} exhibits four more noisy cases (all models are sampled with noise level $\sigma=0.5\%$). Compared with other methods, ours achieves the best balance between preserving the overall topological structure and surface details. Table~\ref{tbl:eval_noise_both} reports the quantitative results for the models in Fig.~\ref{fig:noise} and Fig.~\ref{fig:noise2}.

\subsubsection{Real-world Scans}

Different from synthetic data, real-world scans generally suffer from several kinds of imperfections, such as non-uniform point distribution, scanner noise, outlier points, multi-view misalignment, occluded missing regions, etc. Furthermore, scanner data are usually large in scale, having millions of points. To evaluate the practicability of our method on scanned data, we experiment with three raw range scans (Armadillo, Bunny and Dragon) from the Stanford scanning repository~\citep{curless1996vmbcmri,krishnamurthy1996,turk1994,gardner2003} and one scan (Lady) from \citet{lu2018gr}. Note that some raw scans contain background points as well, and we manually set a bounding box to filter out background points.

Fig.~\ref{fig:scans} shows the reconstruction results using normals estimated by our method, iPSR~\citep{hou2022ipsr} and Dipole~\citep{metzer2021dipole}. To adapt to scanner noise, we set $w_1=0.01$ and $w_2=0.04$ for our method, and set the point weight to 0.5 for iPSR and the final reconstruction step. Dipole still suffers from severe misorientation (see the Armadillo head and the Dragon tail). Even though iPSR can infer a mostly correct orientation, it is less capable of handling scanner noise. Our method not only produces globally consistent orientations but also achieve a good balance between noise handling and detail preserving.

\subsection{Convergence Analysis}\label{sec:convergence-analysis}
In this section we compare the convergence processes of our method and the baselines. In Fig.~\ref{fig:itertest_all} and Fig.~\ref{fig:itertest_quanti}, we present the qualitative and the quantitative results for the intermediate convergence steps using two models: 3DBenchy with 8,000 points and XYZRGB Statuette with 50,000 points. For fairness, we show the first 40 iterations for all iterative methods. Note that our method is able to achieve an overall correct orientation in the first few iterations. Quantitatively, our method also achieves lower normal error than other baselines after convergence.

\subsection{Complexity Analysis}

In this section we study one important aspect of our method: the performance improvement in terms of space/time complexity. Considering the complexity of the baselines, we divide the experiments into three groups: small-scale (1,000$\sim$5,000 points), medium-scale (10,000$\sim$50,000 points) and large-scale (100,000$\sim$1,000,000 points). For the small-scale study, we experiment with all methods. We exclude PGR from large-scale experiments, and exclude GCNO from both medium-scale and large-scale experiments. For each experiment, we use an Armadillo model uniformly sampled without noise.

Furthermore, to present a more comprehensive analysis, we break down each method into a preprocessing stage and a main stage. For our method, iPSR and PGR, the preprocessing stage includes octree building, and the main stage includes normal solving. For GCNO, preprocessing includes Delaunay triangulation and the main stage includes the LBFG-S iterations. For Dipole, preprocessing includes all steps prior to inter-patch propagation, and the main stage includes everything else. Note that the final reconstruction stage and file I/O are excluded for all methods.

\begin{figure*}
  \centering
  \begin{picture}(\linewidth,0.5cm)
  \put(0.08\linewidth,0.1cm){\small $T_{\rm pre}+T_{\rm main}$ (seconds)}
  \put(0.33\linewidth,0.1cm){\small $T_{\rm pre}$ (seconds)}
  \put(0.57\linewidth,0.1cm){\small $T_{\rm main}$ (seconds)}
  \put(0.83\linewidth,0.1cm){\small Memory (MB)}
  \end{picture}
  \includegraphics[width=\linewidth]{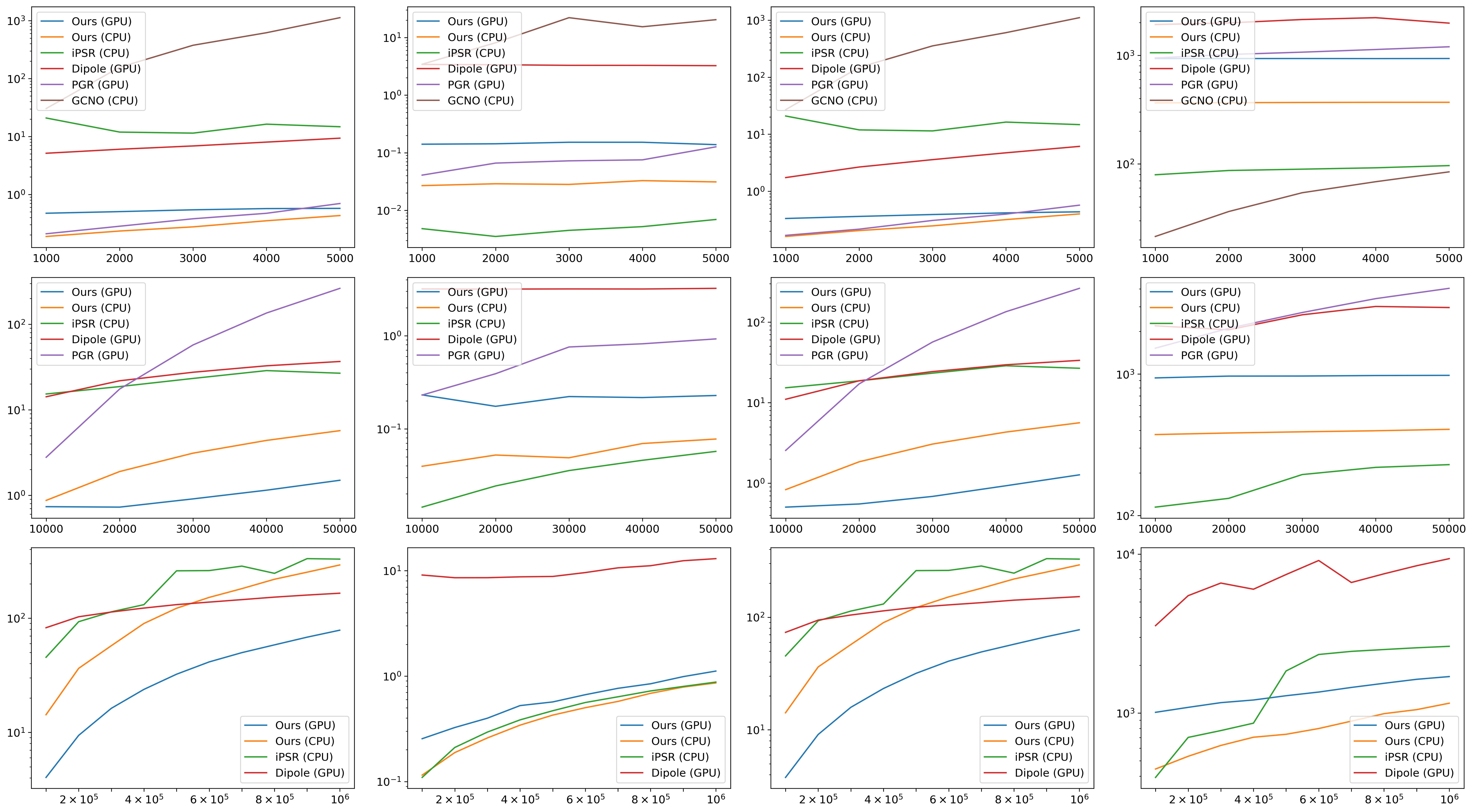}
  \caption{Complexity comparisons of different methods, evaluated on a uniformly sampled Armadillo model with different sample numbers (first row: small-scale; second row: medium-scale, third row: large-scale). We show the total running time $T_{\rm pre}+T_{\rm main}$, preprocessing time $T_{\rm pre}$, main-stage execution time $T_{\rm main}$ and peak memory usage (MB).\label{fig:complexity}}
  \Description{Complexity analysis of different methods.}
\end{figure*}

In Fig.~\ref{fig:complexity}, we show the total running time $T_{\rm total}=T_{\rm pre}+T_{\rm main}$, preprocessing time $T_{\rm pre}$, main stage running time $T_{\rm main}$ and the total peak memory (CPU plus GPU). Time and memory measurements are shown in log-scale due to large variations between different methods. In terms of total running time, it can be seen that our method is the fastest for medium and large inputs, and even our CPU implementation with OpenMP is faster than other methods. Note that for small-scale inputs the CPU implementation is faster due to the overhead associated to CUDA runtime library and cross-device data transfer. In terms of memory consumption, our method is slightly larger than iPSR and GCNO for small and medium inputs, mainly because our implementation involves PyTorch and CUDA overheads. However, as the input size scales up, our method has the lowest memory footprint.

Additionally, we would like to highlight our time and memory consumption improvements over the other two winding number-based methods: PGR and GCNO. Both of PGR and GCNO requires repeated evaluations of winding numbers, and both do so in a naive way by directly summing all contributions from each input point, leading to an $N^2$ complexity lower-bound for each evaluation ($N$ being the number of points). More specifically, to orient a point cloud with 100,000 points, PGR requires at least 3000 GB memory and GCNO would take weeks. On the other hand, our treecode-based algorithm is known to have $O(N\log N)$ complexity~\citep{barnes1986treecode}. Thus, our method easily scales up to millions of points and can finish normal orientation for these large-scale inputs in minutes.

\subsection{Evaluation of Individual Components}

In this section, we evaluate individual components of our algorithm: the smoothing width in Sec.~\ref{subsec:width} and the iterative steps in Sec.~\ref{subsec:iterative-algorithm}.

\subsubsection{Smoothing width}\label{subsubsec:width}
\begin{figure}
  \centering
  \includegraphics[width=\linewidth]{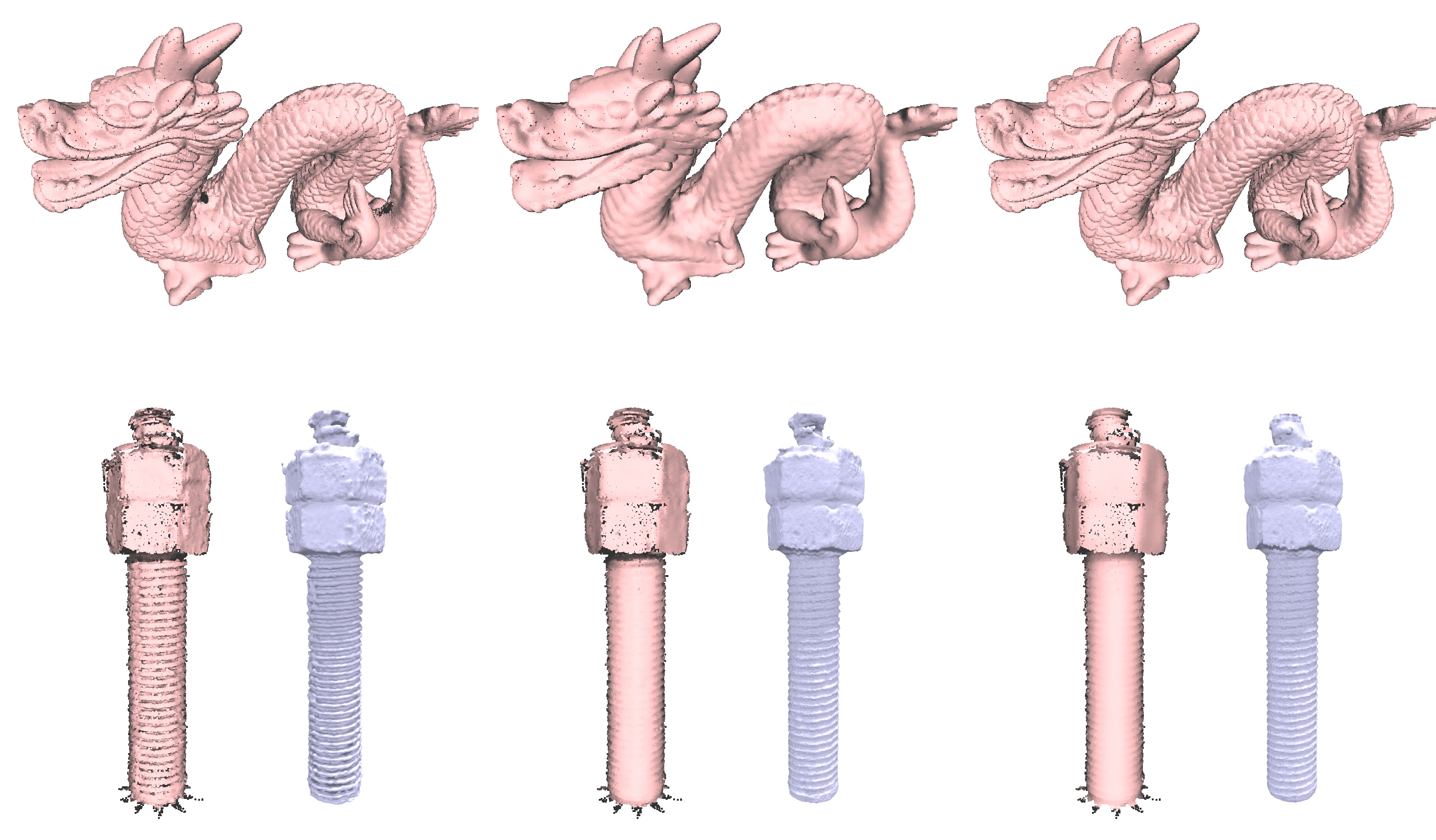}
  \begin{picture}(\linewidth,0.3cm)
    \put(0.04\linewidth,3.3cm){\small $w_1=w_2=0.002$}
    \put(0.38\linewidth,3.3cm){\small $w_1=w_2=0.016$}
    \put(0.69\linewidth,3.3cm){\small $w_1=0.002,\ w_2=0.016$}
    \put(0.02\linewidth,0.2cm){\small $w_1=0.01,\ w_2=0.04$}
    \put(0.36\linewidth,0.2cm){\small $w_1=0.02,\ w_2=0.08$}
    \put(0.7\linewidth,0.2cm){\small $w_1=0.04,\ w_2=0.16$}
  \end{picture}
  \caption{Study of the effect of the smoothing width. First row: decreasing the width during the iterative process allows both consistent orientation and detail preserving. Second row: adjusting widths allows explicit control over the smoothness of normals.\label{fig:width_study}}
  \Description{}
\end{figure}

As mentioned in previous works~\citep{lu2018gr,lin2023pgr} that also apply smoothing modifications to the winding number formula, this modification allows ignoring local points, leading to better robustness to noise or non-uniformity. Intuitively, a large smoothing width produces a more consistent orientation but may result in oversmoothing, and a small smoothing width preserves more details but may fail to orient normals consistently. In contrast to using an invariant smoothing width as in prior work~\citep{lin2023pgr}, our width scheduling scheme that starts at a large width and ends with a small width achieves both consistent orientation and detail preserving. The first row in Fig.~\ref{fig:width_study} compares using invariant smoothing widths and decreasing smoothing widths. As is evident from Fig.~\ref{fig:width_study}, a small width leads to orientation errors in some concave regions while a large width causes oversmoothing. Our scheduling scheme achieve consistent and detailed normal estimation.

The smoothing width can also be tuned to adjust normal smoothness at the user's will. The second row in Fig.~\ref{fig:width_study} shows three different smoothing width configurations for processing a noisy scan. A larger smoothing width leads to a smoother result.

We remark that we also tried other scheduling schemes (e.g., cosine, exponential) but found no obvious difference, as long as the width drops from a large $w_2$ to a small $w_1$ smoothly. Thus, we chose the most simple linear scheduling. Furthermore, due to the width scheduling scheme, we do not adopt any early termination strategy because it may stop at a large $w$, resulting in oversmoothed normals. Note that using a fixed number of iterations has little practical impact, because running full 40 iterations with our method is still faster than 1 iteration of the baselines.

\begin{figure}
  \centering
  \includegraphics[width=0.7\linewidth]{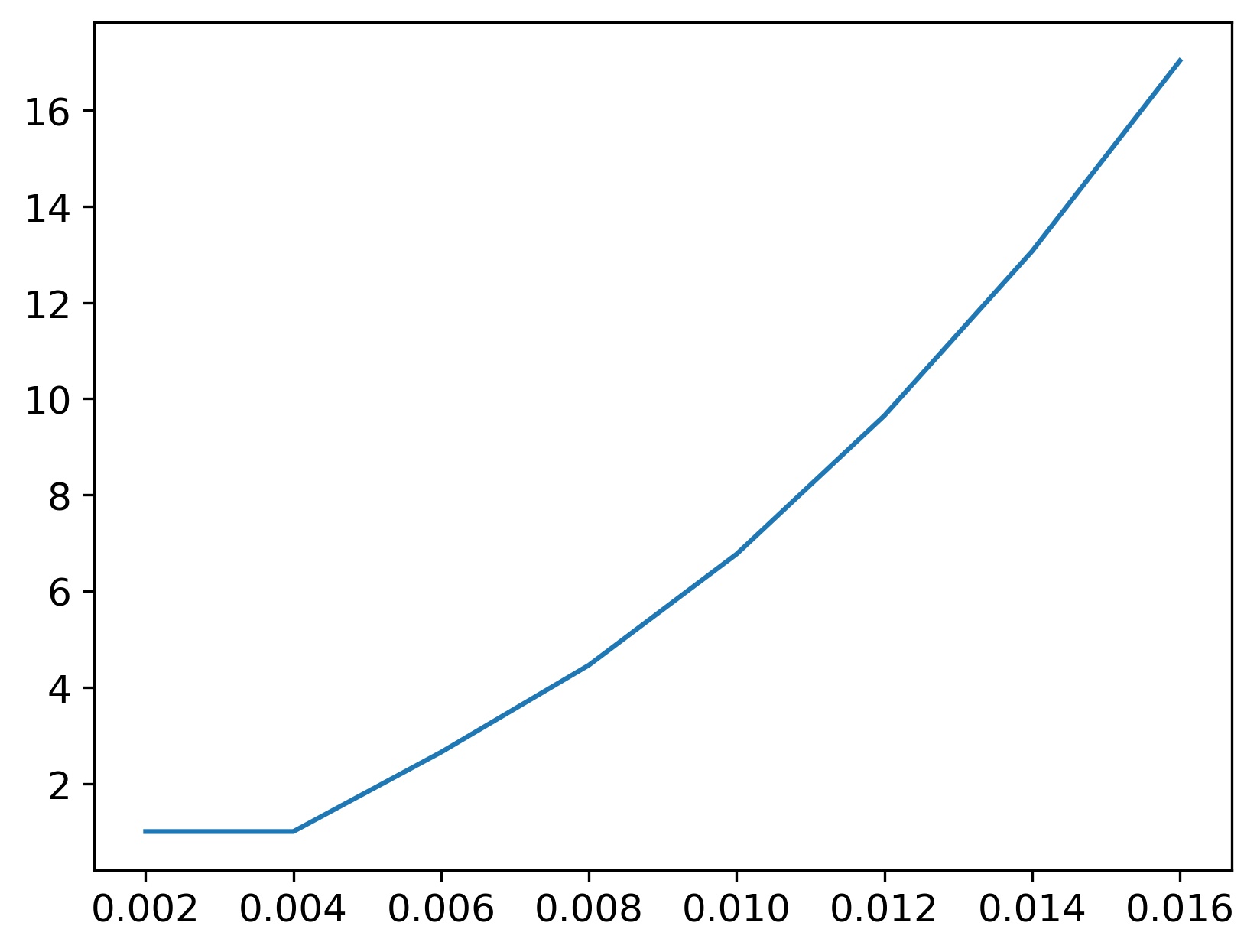}
  \begin{picture}(\linewidth,0.2cm)
  \put(0.5\linewidth,0.2cm){\small $w$}

  \put(0.04\linewidth,2.6cm){\small $N_{\rm nbr}$}

  \end{picture}
  \caption{Plot of the average number $N_{\rm nbr}$ of neighbor points in the $w$-neighborhood of each point (including itself) in a uniformly sampled Armadillo model with 100,000 points. The average point spacing is $0.0076$. All the above statistics are computed after normalizing to the unit cube.\label{fig:n_exclude}}
  \Description{Plot of the average number of neighbors points in a $w$-neighborhood.}
\end{figure}

To provide a more intuitive view of the smoothing width, we report the average number of neighbors in the $w$-neighborhood of a 100,000-point uniformly sampled Armadillo model in Fig.~\ref{fig:n_exclude}.

\subsubsection{Steps in the Iterative Algorithm}
We discuss the necessity of different steps in Algorithm~\ref{alg:wnncalg-final}. In fact, the necessity of the gradient step is easy to see: without it the algorithm simply stays at $\mu=0$ across the whole process. The rescaling step is also necessary because the scale change before and after the WNNC update is typically $10^4\sim10^{10}$, depending on point density. Without rescaling the normals would soon blow up numerically.

\begin{figure}
  \centering
  \includegraphics[width=\linewidth]{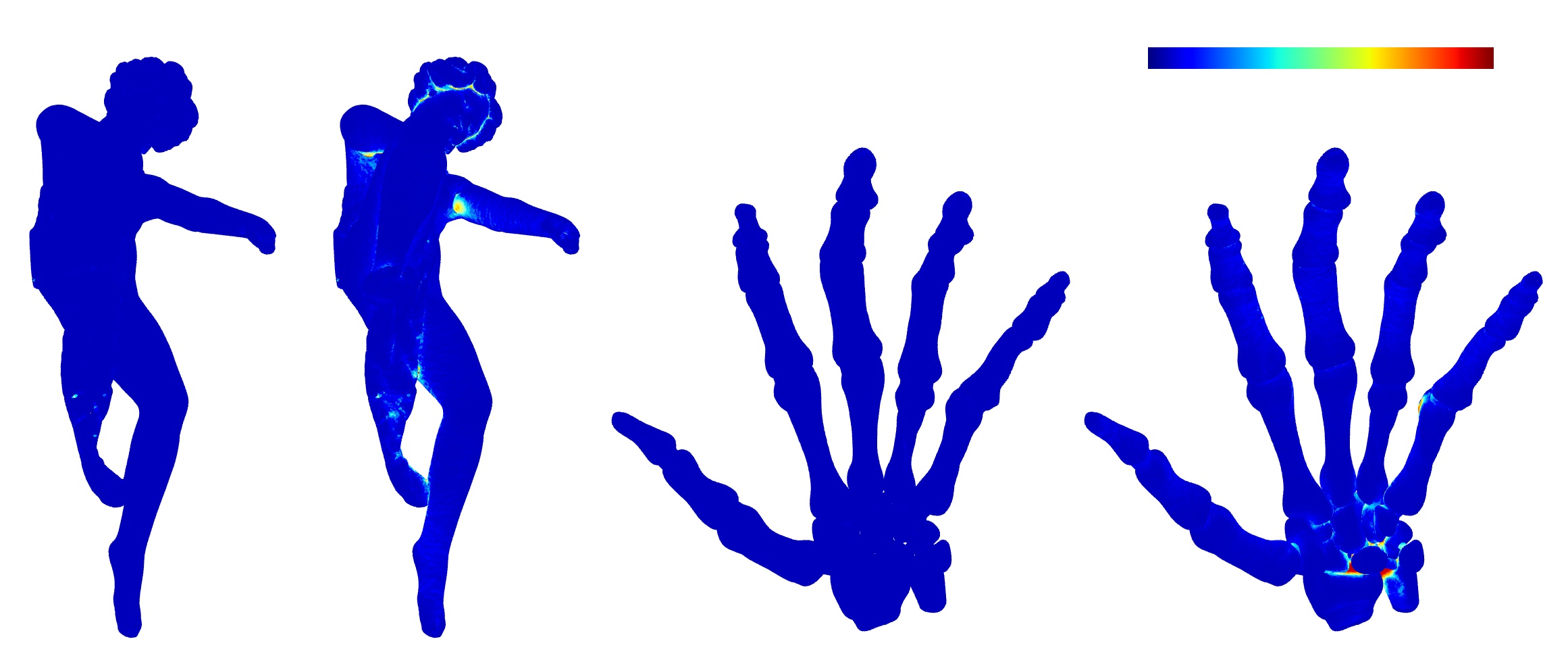}
  \begin{picture}(\linewidth,0.3cm)
  \put(0.06\linewidth,0.2cm){\small w/}
  \put(0.25\linewidth,0.2cm){\small w/o}
  \put(0.54\linewidth,0.2cm){\small w/}
  \put(0.83\linewidth,0.2cm){\small w/o}
  \put(0.70\linewidth,3.58cm){\small 0}
  \put(0.97\linewidth,3.58cm){\small 1}
  \end{picture}
  \caption{Ablation study of the effectiveness of WNNC updates. We show angular error maps of normals estimated with (w/) and without (w/o) using the WNNC updates. The latter introduces notably larger angular error especially in concave regions.\label{fig:abl_nownnc}}
  \Description{Ablation study of the WNNC updates.}
\end{figure}

It remains to examine the effectiveness of the WNNC updates $\mu\mapsto G(\mu)$. In Fig.~\ref{fig:abl_nownnc} we show error maps of results obtained with and without the WNNC updates. These results suggest that estimated normals estimated without the WNNC updates would be highly inaccurate, especially in concave regions.

\section{Discussion}

\subsection{Extended Discussions}
In this section we conduct extended discussions to provide more insights into our method.

\subsubsection{Why not use estimated local areas during optimization?}\label{sec:uselocalareas}
\begin{figure}
  \centering
  \includegraphics[width=\linewidth]{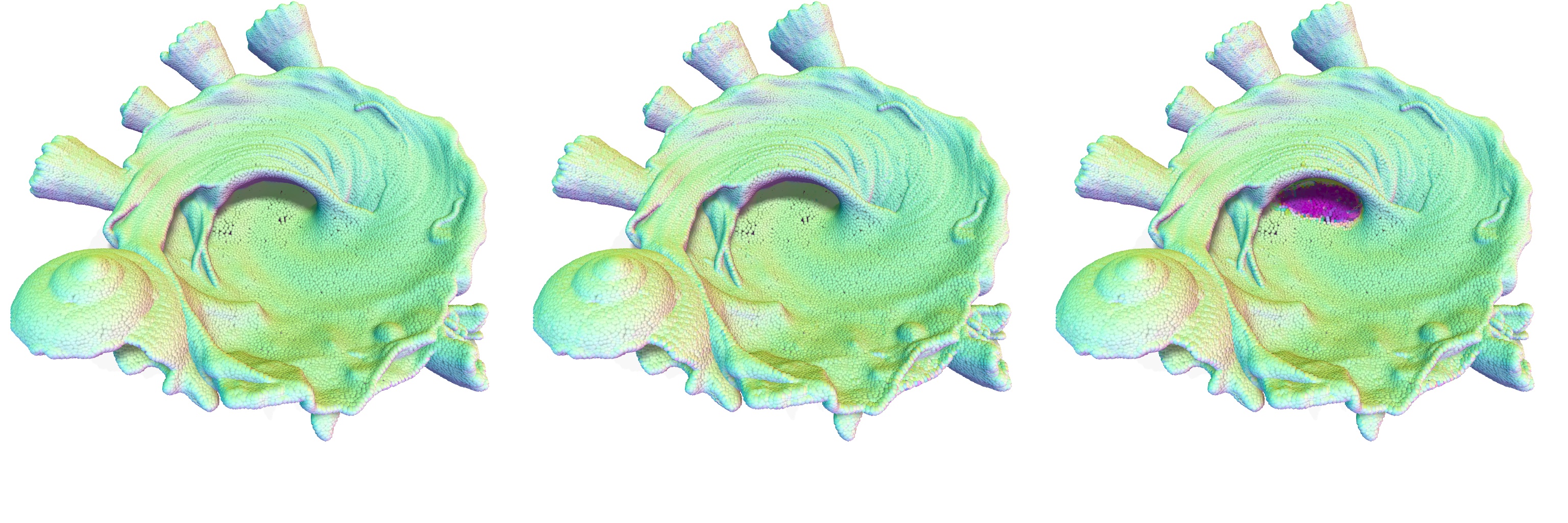}
  \begin{picture}(\linewidth,0.0cm)
  \put(0.08\linewidth,0.2cm){\small GT normals}
  \put(0.39\linewidth,0.2cm){\small Original method}
  \put(0.72\linewidth,0.3cm){\small Using estimated $\sigma_i$}
  \put(0.77\linewidth,0.0cm){\small and Eq.~\eqref{eq:usegtarea}}

  \end{picture}
  \caption{We test the alternative where each local surface element is enforced to have a pre-estimated area. Experiments show this will lead to notable orientation errors (the cave-like part in the middle). Points rendered here are colored by their normals.\label{fig:areatest-usegtarea}}
  \Description{Experiments of whether our method benefits from estimated surface areas.}
\end{figure}

Using estimated local surface element areas $\sigma_i$ adds constraints $|\mu_i|=\sigma_i$ to Eq.~\eqref{eq:value-energy}, making the feasible space of $\mu$ non-linear. Consequently, Algorithm~\ref{alg:gradstep} must be replaced by a non-linear solver, which is not immediately or easily possible using our operators. As an alternative, assuming known/estimated local areas $\sigma_i$, we can enforce $|\mu_i|=\sigma_i$ by changing the rescaling step in Algorithm~\ref{alg:wnncalg-final} to
\begin{equation}
\mu_i\leftarrow\hat\mu_i(\sigma_i/|\hat\mu_i|).\label{eq:usegtarea}
\end{equation}
In Fig.~\ref{fig:areatest-usegtarea}, we use the GT mesh to compute Voronoi areas, and proceed with Eq.~\eqref{eq:usegtarea}. As Fig.~\ref{fig:areatest-usegtarea} shows, using fixed local areas leads to notable orientation errors. This suggests even with accurate local areas, our algorithm cannot benefit from using fixed areas because the feasible space becomes non-linear. Furthermore, in practice, local area estimation can be difficult or inaccurate especially for noisy inputs or sharp features.
In summary, using fixed estimated areas makes the solution space non-linear, and makes Eq.~\eqref{eq:value-energy} harder to solve, is less robust to noise, leads to extra computational complexity and brings no obvious benefit. Hence, we do not use estimated local areas.

\subsubsection{Do solved local areas converge to actual local areas?}
Not necessarily. In Table~\ref{tbl:sphere_area_cvg}, we conduct an experiment on the area convergence using a unit sphere model. We take its vertices as the input point cloud (163,842 points), whose local Voronoi areas $\{\sigma_i^{\rm GT}\}$ are directly computed from the GT triangulation. We empirically find after 40 iterations when the normals are already accurate, solved local areas $\{|\mu_i|\}$ do not accurately match GT areas $\{\sigma_i^{\rm GT}\}$, while the $k$NN-based area estimation method~\citep{barill2018fastwind} is more accurate, as shown in Table~\ref{tbl:sphere_area_cvg}. We remark that this work does not claim area convergence, and whether areas converge or not is irrelevant, because area convergence is not required for accurate normal convergence.

\begin{table}
    \centering
    \caption{A comparison of the statistics for GT surface element Voronoi areas $\{\sigma_i^{\rm GT}\}$, estimated areas $\{\sigma_i^{k\rm NN}\}$ using $k$ nearest neighbors as in \citet{barill2018fastwind}, and solved areas $\{|\mu_i|\}$ for a uniformly sampled unit sphere model.\label{tbl:sphere_area_cvg}}
    \begin{tabular}{c|c|c|c}
        \toprule
        & $\{\sigma_i^{\rm GT}\}$ & $\{\sigma_i^{k\rm NN}\}$ & $\{|\mu_i|\}$ \\
        \midrule
        Min & $6.795\times10^{-5}$ & $6.794\times10^{-5}$ & $7.465\times10^{-5}$ \\
        Max & $9.253\times10^{-5}$ & $9.252\times10^{-5}$ & $8.330\times10^{-5}$ \\
        Mean & $7.670\times10^{-5}$ & $7.670\times10^{-5}$ & $7.858\times10^{-5}$ \\
        Total & $12.566$ & $12.566$ & $12.875$ \\
        \bottomrule
    \end{tabular}
\end{table}

\begin{figure}
  \centering
  \includegraphics[width=0.9\linewidth]{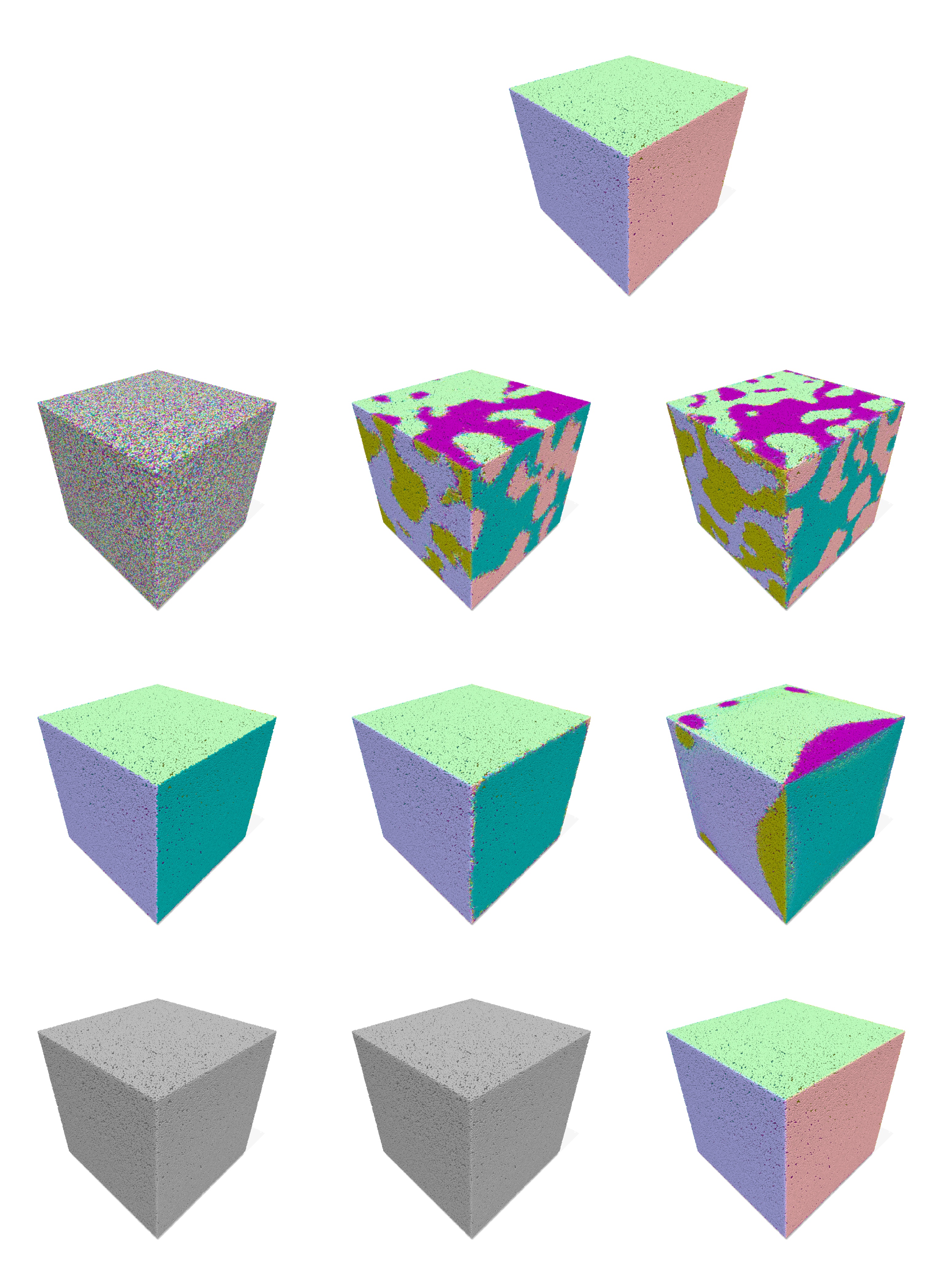}
  \begin{picture}(\linewidth,0.2cm)
  \put(0.09\linewidth,0.2cm){\small Initial normals}
  \put(0.41\linewidth,0.2cm){\small w/o grad. step}
  \put(0.72\linewidth,0.2cm){\small w/ grad. step}

  \put(0.04\linewidth,2.8cm){\small Zero init.}
  \put(0.04\linewidth,5.4cm){\small One side flipped}
  \put(0.04\linewidth,7.9cm){\small Random init.}

  \put(0.35\linewidth,9.6cm){\small GT normals}

  \end{picture}
  \caption{Normal orientation results obtained by different normal initialization strategies and whether the gradient step (Algorithm~\ref{alg:gradstep}) is used.\label{fig:fliptest}}
  \Description{Normal orientation results obtained by different normal initialization strategies and whether the gradient step is used.}
\end{figure}

\begin{figure}
  \centering
  \includegraphics[width=\linewidth]{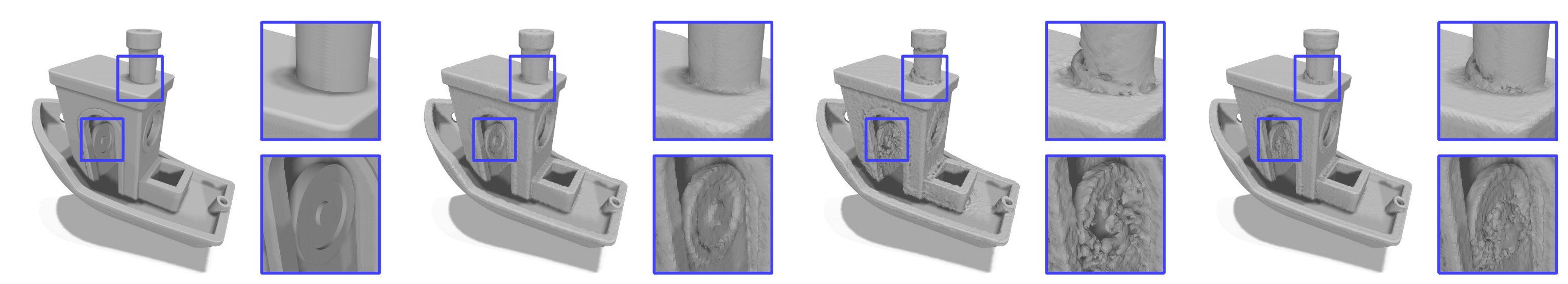}
  \begin{picture}(\linewidth,0.5cm)
  \put(0.09\linewidth,0.2cm){\small GT mesh}
  \put(0.38\linewidth,0.2cm){\small Ours}
  \put(0.55\linewidth,0.35cm){\small PGR normals}
  \put(0.59\linewidth,0.05cm){\small +SPSR}
  \put(0.8\linewidth,0.35cm){\small PGR normals}
  \put(0.84\linewidth,0.05cm){\small +WNF}

  \end{picture}
  \caption{Comparisons between our method and PGR using different reconstruction methods. Our method outperforms PGR regardless of whether it uses Screened Poisson Surface Reconstruction (SPSR) or the winding number field-based reconstruction (WNF).\label{fig:fair-cr}}
  \Description{Comparisons with PGR in a more fair setting.}
\end{figure}

\subsubsection{Can WNNC alone converge to consistent normals?}\label{sec:wnncalone}
No. Note that a consistently outward normal vector field and a consistently inward vector field are both invariant up to scaling under WNNC updates, which suggests WNNC updates alone cannot handle the inside/outside ambiguity. Furthermore, we empirically found two key factors in the convergence of our algorithm: zero initialization and the gradient steps w.r.t. Eq.~\eqref{eq:value-energy}. In Fig.~\ref{fig:fliptest}, we experiment with different initialization strategies and whether Eq.~\eqref{eq:value-energy} is applied. As Fig.~\ref{fig:fliptest} shows, if the normals are not zero-initialized, the normals cannot converge to a consistent orientation regardless of whether Eq.~\eqref{eq:value-energy} is used. On other hand, if we only apply WNNC updates to zero-initialized normals without Eq.~\eqref{eq:value-energy}, the optimization would simply stay at zero vectors according to Algorithm~\ref{alg:wnncalg-final}. Good convergence is only possible with both zero initialization and Eq.~\eqref{eq:value-energy}.

\subsubsection{WNNC vs. PGR}
While the effectiveness of our method seems to heavily rely on the objective Eq.~\eqref{eq:value-energy} proposed by PGR~\citep{lin2023pgr}, the latter is fundamentally built on an underdetermined system (Eq.~\eqref{eq:value-energy}). As discussed in \citet{lin2023pgr}, even if given enough iterations, their solved normals usually do not align well with GT normals, resulting in low normal accuracy. On the other hand, our WNNC formulation provides a new set of constraints for normal directions, leading to a more determined system. By alternating gradient steps w.r.t. Eq.~\eqref{eq:value-energy} and WNNC updates, our method quickly converges to an accurate and consistently oriented normal vector field. As shown in Sec.~\ref{sec:convergence-analysis}, our method achieves much higher accuracy given the same number of iterations. Other comparisons also show our method achieves better accuracy after full convergence.

Furthermore, the authors of PGR~\citep{lin2023pgr} suggested that the normals solved by PGR are not really compatible with SPSR~\citep{kazhdan2013spr}, and better reconstructions can be obtained by plugging these normals back into the winding number field (WNF). While we have previously used SPSR as the reconstruction for all methods for fairness, we also test PGR with WNF-based reconstruction as in \citet{lin2023pgr}. In Fig.~\ref{fig:fair-cr}, we observe that our method still outperforms PGR despite the final reconstruction method.

\subsection{Conclusion}
In this work, we propose a novel property, namely the winding number normal consistency (WNNC), derived from the winding number formula to address the problem of estimating globally consistent normal orientations. We turn the WNNC property into a practical, efficient yet embarrassingly simple iterative algorithm. We also observe the special structures in the iterative algorithm that allows acceleration using the treecode algorithm~\citep{barnes1986treecode}. Lacking an existing package of the treecode algorithm that suits our needs (GPU-based and supporting our operators), we implement our own CUDA kernels. With extensive experiments on a variety of datasets, we exhibit the superior performance of our algorithm and implementation over recent state-of-the-art methods, in terms of both normal orientation quality and computational costs. Furthermore, our implementation is integrated with PyTorch~\citep{ansel2024pytorch2,paszke2019pytorch} and is easy to use. We hope our implementation can facilitate future research into winding numbers.

\begin{figure}
    \centering
    \includegraphics[width=\linewidth]{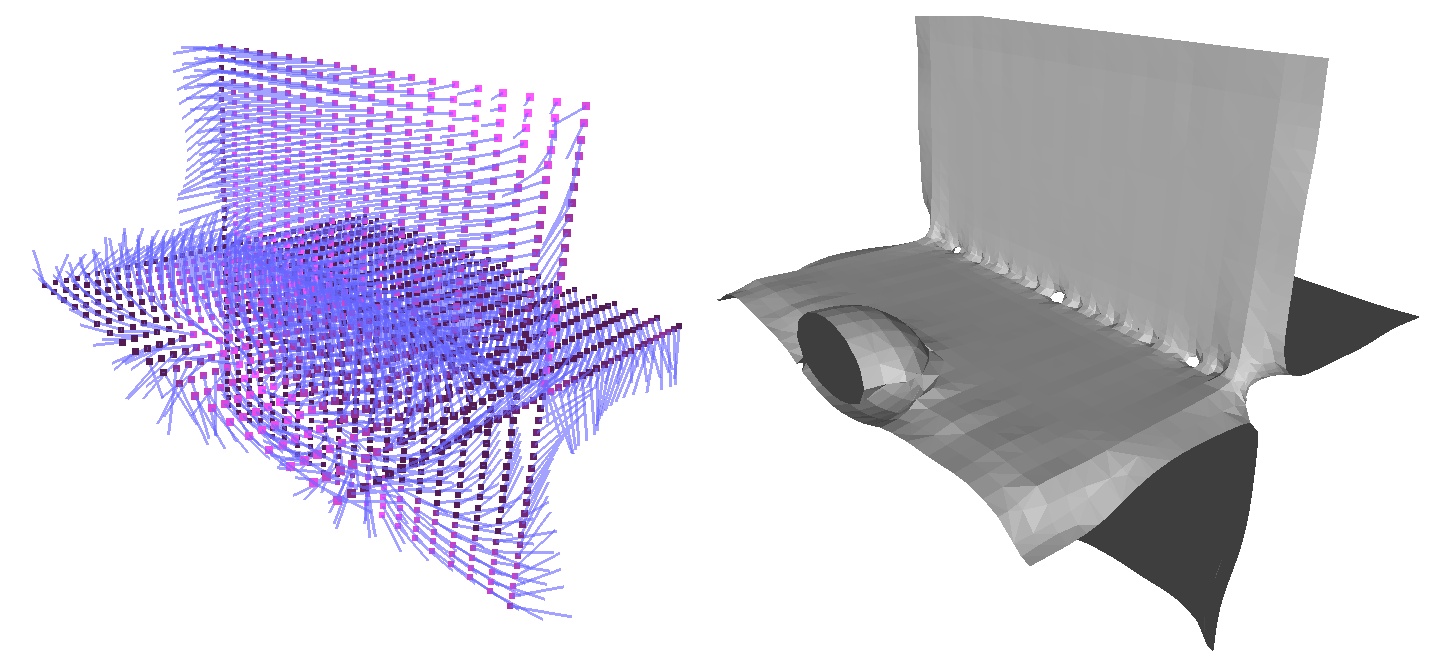}
    \caption{Our method is built on the winding number formula, which cannot handle non-manifold surfaces. \label{fig:cross}}
    \Description{Failure case of open, non-manifold surfaces.}
  \end{figure}

\subsection{Limitations and Future Work}
Despite the good performance of our algorithm, there are still a few aspects that could be improved or are worth exploring.

\paragraph{Smoothing widths}
The normal estimation quality of our method still highly depends on the smoothing width. As explained in Sec.~\ref{subsubsec:width}, an improper choice can lead to either wrong orientation or oversmoothing. We empirically found that the widths can be kept the same for data with similar noise levels and structural complexity. Hence, adjusting the width parameters shouldn't take much effort. However, it is still desirable for future work to explore how to automatically determine the optimal width parameters.

\paragraph{Non-manifold surfaces or open scans}
The underlying mathematical tool of our method, the winding number formula, is meant to describe 3D shapes with a clear partitioning of space into its interior and exterior. This means our method cannot handle 3D shapes that does not satisfy this property, e.g. non-manifold surfaces and open scans. Fig.~\ref{fig:cross} shows a failure case of two planes intersecting each other. This geometry is non-manifold and does not enclose a spatial region. Our method, as well as any other implicit field-based method, is not suitable for these cases. However, as open scans are also very common in the real world, e.g., street scenes, future work should explore how to extend the winding number formula to these cases.

\paragraph{Further efficiency improvement}
Currently, we use the treecode algorithm to accelerate winding number related evaluations because it is easy to implement and parallelize. A possible future direction is to implement a more advanced acceleration technique, the fast multipole method (FMM)~\cite{greengard1997fmm}, which further reduces complexity and allows precision control. However, FMM requires more careful mathematical analysis and is more difficult to implement and parallelize. Thus, we leave it as future work. 

\paragraph{Accelerating existing winding number-based algorithms}
Another line of future work would be to extend our treecode implementation to accelerate other existing methods that use the winding number field, e.g. GCNO~\citep{xu2023gcno}. Furthermore, note that we have implemented the $A^T$ operator, which is exactly what is needed for differentiating through the winding number formula. This means our implementation can also be incorporated into differentiable programming involving winding numbers. Since our implementation has been incorporated with the PyTorch~\citep{ansel2024pytorch2,paszke2019pytorch} framework, differentiable programming with winding numbers can be made quite easy.

\paragraph{A theoretical proof of convergence}
While our proposed method exhibits notable superiority over other baselines, our solution remains an engineering solution without a rigorous theoretical proof of convergence. Future work should explore the theoretical aspects of the WNNC formulation.

\begin{acks}
The work is supported by the National Science Foundation of China under Grant Number~\grantnum{NSFC}{62125107} and~\grantnum{NSFC}{92370125}.
\end{acks}

\bibliographystyle{ACM-Reference-Format}
\bibliography{refs-wnnc}

\end{document}